\documentclass[aps,pre,reprint, amsmath, amssymb,superscriptaddress]{revtex4-1}

\usepackage{bm}
\usepackage[retainorgcmds]{IEEEtrantools}
\usepackage{graphicx}
\usepackage{mathrsfs}
\usepackage{amsmath}
\usepackage{amssymb}
\usepackage{color}
\usepackage{amsfonts}
\usepackage{times,txfonts}
\usepackage{nicefrac}
\usepackage{selinput}
\usepackage{ulem}

\newcommand{\ud}{\,\mathrm{d}}

\DeclareMathOperator{\tr}{tr}

\begin{document}

\title{
Dynamical chaotic phases and constrained quantum dynamics}
\date{\today}
\author{Andr\'e M. Timpanaro}
\email{a.timpanaro@ufabc.edu.br}
\affiliation{Universidade Federal do ABC,  09210-580 Santo Andr\'e, Brazil}
\author{Sascha Wald}
\email{swald@sissa.it}
\affiliation{SISSA - International School for Advanced Studies and INFN, via Bonomea 265, I-34136 Trieste, Italy}
\author{Fernando Semi\~ao}
\affiliation{Universidade Federal do ABC,  09210-580 Santo Andr\'e, Brazil}
\author{Gabriel T. Landi}
\affiliation{Instituto de F\'isica da Universidade de S\~ao Paulo,  05314-970 S\~ao Paulo, Brazil}

\begin{abstract}

In classical mechanics, external constraints on the dynamical variables can be 
easily implemented within the Lagrangian formulation. Conversely,
the extension of this idea to the quantum realm, which dates back to Dirac, has 
proven notoriously difficult due to the non-commutativity of
observables. Motivated by recent progress in the experimental control of quantum 
systems, we propose a framework for the implementation of 
quantum constraints based on the idea of work protocols, which are dynamically 
engineered to enforce the constraints. As a proof 
of principle, we consider the dynamical mean-field approach of the many-body 
quantum spherical model, which takes the form of a quantum
harmonic oscillator plus constraints on the first and second moments of one of 
its quadratures. The constraints of the model are implemented by 
the combination of two work protocols, coupling together the first and second 
moments of the quadrature operators. We find that such constraints 
affect the equations of motion in a highly non-trivial way, inducing non-linear 
behavior and even classical chaos. Interestingly, Gaussianity is 
preserved at all times. A discussion concerning the robustness of this approach 
to possible experimental errors is also presented.

\end{abstract}
\maketitle{}

%
%
%
%

\section{Introduction}
Since the conception of quantum mechanics, the
 implementation of non-trivial dynamical effects that go beyond the linearity of 
Schr\"odinger's equation has been a recurring topic of research.
Recently, this search has seen a renewed interest, particularly due to  
developments in quantum platforms such as ultra-cold atoms 
\cite{Landig2016,Kaufman2016,Bordia2016,Hruby2017}. 
For many-body systems, non-trivial effects such as quantum chaos 
\cite{Gutzwiller1971,Bohigas1984,Haake2001} and criticality  \cite{Sachdev1998} 
emerge naturally from the complexity of the many-body Hilbert space.
\textcolor{black}{It is well-known that - up to leading order - 
such effects of complex many-body interactions can be captured by external 
constraints,
see e.g. large $n$ quantum field theories \cite{Moshe2003} or the spherical 
model \cite{Berlin1952,Lewis1952,Henkel1984,Vojta1996}.}
In these systems, the complexity of strongly interacting 
degrees of freedom 
is reduced to the solution of a single, transcendental equation that stems from 
a certain external constraint.
These types of constraints can be enforced in modern experiments, 
using techniques such as continuous measurements in order to project the dynamics
onto specific subspaces (the Zeno effect), rendering the dynamics effectively
non-linear \cite{Facchi2000,Touzard2017}.

%

Remarkably, even \textit{simple} systems with few degrees of freedom 
can exhibit effects such as bistability and criticality 
in certain limits of the Hamiltonian parameters
\cite{Drummond1999,Casteels2017,Katz2007,Hwang2015,Carmichael2015}.
This motivates the present study in which we shall explore these effects
by means of constraints that act on a free system.

Despite the potential for mimicking many-body effects on a 
mean-field level and 
all of the above mentioned experimental applications,
the development of a general framework for the implementation of constraints 
that act directly on quantum observables has proven to be notoriously difficult.
In classical mechanics, constraints can be implemented in a natural 
way within the Lagrangian formulation.
The extension of this idea to quantum systems can be traced all the way back to 
Dirac \cite{Dirac1958} and has ever since been the subject of several studies
\cite{Gustavsson2009,DeOliveira2014,Grundling1998,Bartlett2003,DaSilva2017,
Bartlett2012,Giampaolo2018}.
However, these approaches are of limited applicability since none of them enjoys
the breadth and reach of the Lagrangian formulation. 

The enormous success of the Lagrangian formulation in classical mechanics often overshadows the fact that external constraints are nothing but time-dependent forces following  specific protocols.
That is, any constrained dynamics can always be viewed as an unconstrained evolution subject to carefully tailored external forces (typically tensions and normal forces) that act to enforce the constraints at all times. 
Of course, within classical mechanics this viewpoint  is not at all necessary. 
Here, we adopt this point of view and show how constraints can arise from carefully chosen work protocols. Other alternatives based on different approaches have also been proposed, what include, for instance, quantum feedback control \cite{Eastman2017}, quantum Zeno effect \cite{Gong2016,Touzard2017}, active entanglement control \cite{Leandro2010}, shortcuts to adiabaticity \cite{Berry2009,Chen2010,DelCampo2014,Abah2017}, dynamical decoupling \cite{Teixeira2016} and constrained quantum annealing \cite{Kudo2018,Hen2016a,Hen2016}. In the context of work protocols, modern developments in quantum thermodynamics use this concept in varied applications such as in the elaboration of quantum heat engines \cite{Kosloff2014,Rosnagel2015,Alicki2006,Alicki2012}.

\textcolor{black}{In this work, we explore this idea further by putting forth a way of implementing quantum constraints, by means of external agents performing engineered protocols. To illustrate this framework in action}, we focus on a quantum harmonic oscillator and discuss how to engineer work protocols that enforce a constraint between the first and second moments of the quadrature operators. 
\textcolor{black}{These constraints shall be motivated by \textit{mean-field} approximation 
schemes to a stereotypical many-body quantum system, namely the spherical model. Thus, the 
present study can also be viewed as a quantitative guide to dynamical properties of such systems.
Particularly, our choice of constraints allows us to include mean-field descriptions of many-body 
Floquet systems that have attracted a large amount of interest in the past decade, see
\cite{Buk2015,Gold2014} and references therein.
We will show that, although the quantum evolution preserves Gaussianity at all times, the 
constraints force the system to behave in a highly anharmonic fashion. In this manner a dynamical 
order - disorder quantum
phase transition is induced by the constraints to the system. Remarkably, periodic constraints
allow for a third phase to exist where the dynamics
is de facto chaotic. Here, classical chaotic motion is forced on the quantum evolution and we 
shall name this phase the chaotic phase.
} 
Finally, we test numerically the robustness of this approach with respect to small perturbations in the preparation of the initial state and on the work protocols, finding that the errors scale linearly with the size of the perturbations and sub-linearly in time.
In view of these results and of the recent advances in the coherent control of quantum systems, particularly in platforms such as trapped ions and superconducting qubits, we believe that this framework could pave the way for the design of more general quantum constraints.

\section{Implementing Quantum Constraints With External Agents}
\label{sec:framework}

To understand the way we will be implementing quantum constraints, we first note that we can think of a quantum constraint as a functional equation for the 
density matrix $\rho$ of a quantum system
\begin{equation}
\phi\left[\rho\right] = f(t)
\label{eq:constraint-general}
\end{equation}
that must be satisfied at all times, for some functional $\phi$. In order to impose this constraint using external agents $\alpha_i(t)$, we can consider a time dependant Hamiltonian $H(t)$, given by
\begin{equation}
H(t) = H_0 + \sum_j \alpha_j(t) H_j
\end{equation}
where $H_0$ is the Hamiltonian of the unconstrained system in absence of external agents. The unitary evolution of this system is then
\begin{equation}
\frac{\partial\rho}{\partial t} = -i[H_0, \rho] 
-i\sum_j \alpha_j(t) [H_j, \rho]
\end{equation}
meaning that the time derivative of the density matrix depends explicitly on the external agents. The central idea in order to specify the work protocol of the external agents
is to consider an initial condition $\rho_0$ that satisfies Eq.~(\ref{eq:constraint-general}) and taking a time derivative in both sides of Eq.~(\ref{eq:constraint-general}):
\begin{equation}
\frac{\partial}{\partial t}\left(\phi\left[\rho\right]\right) = f'(t).
\label{eq:constraint-derivative}
\end{equation}
If the time derivative of $\phi\left[\rho\right]$ depends explicitly on the external agents, we can solve Eq.~(\ref{eq:constraint-derivative}) by choosing the agents conveniently. However if it does not depend on the agents explicitly, we can repeat the procedure, using now Eq.~(\ref{eq:constraint-derivative}) as a new constraint until we can tie all the constraints to the agents.

More constraints could in principle be added by simply repeating the procedure, as long as we have enough external agents to satisfy all of them. This procedure yields a set of equations that couple the agents $\alpha_j(t)$ and 
$\rho$, meaning that the protocols will be dependent on the initial condition $\rho_0$ of the density matrix.

In order to illustrate this scheme, we apply it to a quantum harmonic oscillator in the next section, using
\begin{align}
\phi_1\left[\rho\right] &\equiv \langle q^2 \rangle \quad\; \; f_1(t) = \lambda = 1
\end{align}
Here $\langle\; \cdot \; \rangle \equiv \tr(\rho\; \cdot\;)$ refers to the average over the time dependent density matrix. The choice of this constraint is motivated by the many-body quantum spherical model, as we shall point out below. Additionally, this 
constraint is advantageous since it only depends on averages, which allows us to formulate the full dynamics in terms of averages only.
%


\section{The Model}
We consider the dynamics of a single bosonic mode, characterized by quadrature operators $q$ and $p$, satisfying the canonical commutation relation
$[q,p]=i$, and subject to the time-dependent Hamiltonian 
\begin{equation}\label{H}
H(t) = \frac{p^2}{2m} + \frac{\mu_t q^2}{2} - B_t q,
\end{equation}
where $m$ is a time-independent constant and $\mu_t$ and $B_t$ are   time-dependent functions, acting as work agents. We shall then focus on implementing the following constraint
\begin{equation}
\label{constraint_1}
\langle q^2 \rangle = \lambda
\end{equation}
in a way that couples the first and second moments of quadrature. Here 
$\lambda$ is a constant that sets the units of the quadrature operators and we shall henceforth set $\lambda =1$.

\subsection{Implementing the constraint}
\textcolor{black}{First, we rephrase the problem in the language of section \ref{sec:framework}.
The Hamiltonian contributions read
\begin{equation}
H_0 = \frac{p^2}{2m},\quad H_1 = \frac{q^2}{2},\quad  H_2 = -q
\end{equation}
whereas with $H_1$ and $H_2$ the work agents
\begin{equation}
\alpha_1(t) = \mu_t,\quad	\alpha_2(t) = B_t,
\end{equation}
are associated in order to fulfill the external constraint
\begin{equation}
\phi\left[\rho\right] = \langle q^2 \rangle,\quad f(t) = 1 
\end{equation}
}
%
%
%
The unitary evolution according to the Hamiltonian~(\ref{H}) yields for the first moments
\begin{IEEEeqnarray}{rCl}
\label{eq_q}
\frac{\ud \langle q \rangle}{\ud t} &=& \frac{\langle p \rangle}{m}, 		\\[0.2cm]
\label{eq_p}
\frac{\ud \langle p \rangle}{\ud t} &=& B_t - \mu_t \langle q \rangle,	
\end{IEEEeqnarray}
whereas the second moments obey
\begin{IEEEeqnarray}{rCl}
\label{eq_q2}
\frac{\ud \langle q^2 \rangle}{\ud t} &=& \frac{2 Z}{m}, 		\\[0.2cm]
\label{eq_p2}
\frac{\ud \langle p^2 \rangle}{\ud t} &=& 2B_t\langle p \rangle - 2\mu_t Z, 	\\[0.2cm]
\label{eq_Z}
\frac{\ud Z}{\ud t} &=& B_t \langle q \rangle + \frac{ \langle p^2 \rangle}{m} - \mu_t \langle q^2 \rangle,
\end{IEEEeqnarray}
with $Z = \frac{1}{2}\langle qp + pq\rangle$.

Following the framework in section \ref{sec:framework}, we must use initial conditions such that $\langle q^2\rangle_0 = 1$ and external agents such that
\begin{equation}
\frac{\partial}{\partial t}\left(\phi\left[\rho\right]\right) = f'(t) \Leftrightarrow \frac{\ud \langle q^2 \rangle}{\ud t} = 0 \Leftrightarrow Z = 0
\label{eq:cond1}
\end{equation}

The condition in Eq.~(\ref{eq:cond1}) does not depend explicitly on $\mu_t$ or $B_t$, so we must use $Z = 0$ as a new constraint. This means that our initial conditions must also be such that $Z_0 = 0$ and repeating what we did for the main constraint we get the condition
\begin{equation}
\frac{\ud Z}{\ud t} = 0 \Leftrightarrow B_t \langle q \rangle +  \frac{\langle p^2 \rangle}{m} -  \mu_t=0
\end{equation}
that can be solved by choosing $\mu_t$ and $B_t$ such that
\begin{equation}
\mu_t = B_t \langle q \rangle +  \frac{\langle p^2 \rangle}{m}
\label{eq:mu-basic}
\end{equation}

\subsection{Choice of the work protocols}

Since we have two work agents, there is some freedom in how we can choose $\mu_t$ and $B_t$. We will use

\begin{equation}
\label{constraint_2}
B_t = \kappa (\langle q \rangle + h \sin \omega t),
\end{equation}
where $\kappa = 1$ is a constant setting the energy scales involved, while $\mu_t$ is defined by Eq.~(\ref{eq:mu-basic}). The motivation for this choice is that Eqs (\ref{H}), (\ref{constraint_1}) and (\ref{constraint_2}) are a generalisation of a system that was previously studied with a zero temperature dissipative Lindblad dynamics \cite{Wald2015b}. It can be viewed as 
a dynamical mean-field study of the quantum spherical model \cite{Henkel1984,Ober72,Vojta1996,Wald2015}
in the following sense. The Hamiltonian of the quantum spherical model with nearest 
neighbour interactions on a hypercubic $d$ dimensional lattice is given by \cite{Ober72,Vojta1996,Henkel1984}
\begin{align}
 \label{sm}
 H_{sm}  = &\sum_{n=1}^{N^d}\big[
 g p_n^2 + \frac{\mu}{2} q_n^2 -Bq_n - J\!\sum_{m\,\in\,\Gamma(n)}\! q_nq_m\big]\\
 &\sum_{n=1}^{N^d}\big< q_n^2\big> = N^d
\end{align}
where $[q_n,p_m] = i\delta_{nm}$ represent bosonic degrees of freedom, $J$ is the nearest neighbour
interaction constant, $\Gamma(n)$ represents the set of nearest neighbours, $\mu$ is a Lagrange
multiplier to ensure the constraint and $g$ quantifies the quantum fluctuations in the system.
The quantum spherical model is the large $n$ limit of the $O(n)$ non-linear $\sigma$ model \cite{Vojta1996} which describes quantum rotors. It
is known to be of great use in order to craft analytical insight into 
statistical properties of many-body systems since it can be exactly solved and shows 
a quantum phase transition in terms of $g$ that is distinct from the mean-field universality class for spatial dimensions $1<d<3$ \cite{Vojta1996,Wald2015}. All 
interactions that go beyond a simple free model are introduced via the highly non-local
constraint which is assured by the Lagrange multiplier $\mu$. In a mean-field approximation
analogous to the Weiss theory of magnetism \cite{Yeo92}, the system (\ref{sm}) can be reduced to
a single body problem $(q,p)$ with a self-consistently determined magnetisation $M = \big<q\big>$. In this scenario Eq. (\ref{sm}) reduces to
\begin{equation} \label{mfsm}
 H_{sm}  = g p^2 + \frac{\mu}{2} q^2 \; ; \; \big< q^2\big> = 1\; ; \;
 \big<q\big> = B
\end{equation}
This system coincides with the one that we are proposing for $h=0$. It has been 
shown that already the system with $h=0$ can lead to surprisingly rich physics
such as non-linear dynamics and quantum freezing by heating effects \cite{Wald2015b}.
In this sense our proposal for $h\neq0$ can be understood as an exploration of 
periodically driven many-body systems in the mean-field regime.

Such studies have seen a large amount of theoretical and experimental
interest \cite{Gold2014,Buk2015,Mossner2017,Kennes2018,Titum15,Kolo2018}.
Routinely, these effects are hard to engineer and even harder 
to theoretically describe. We therefore suggest a systematic exploration of this
simple toy model in order to provide theoretical insight and to see what sort of
dynamical many-body effects can be captured by it. Before we engage in this 
study it is instructive to quantify the effects of the constraint (\ref{constraint_1}) and the choice (\ref{constraint_2}).

\begin{figure}[htbp!]
\centering
\includegraphics[width=0.4\textwidth]{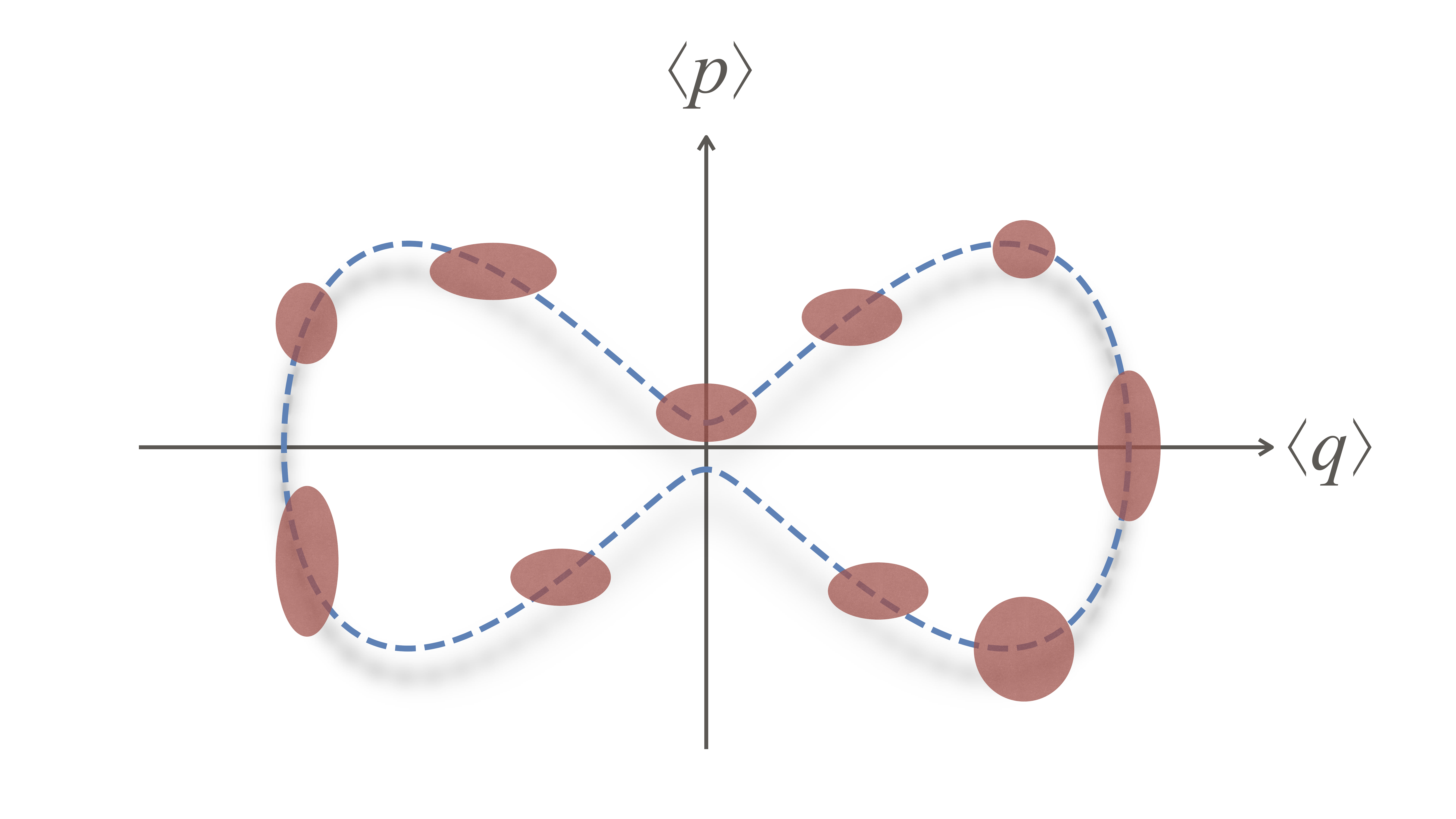}
\caption{\label{fig:diagram}
Diagram of the constrained quantum dynamics. 
Two work agents force a harmonic oscillator to evolve subject to the constraints~(\ref{constraint_1}) and (\ref{constraint_2}). 
This causes the evolution of the first moments $\langle q \rangle$ and $\langle p \rangle$ to couple  to the variances $\langle q^2\rangle - \langle q \rangle^2$ and $\langle p^2\rangle -\langle p \rangle^2$, which have to expand and squeeze depending on the position in phase space.  
The figure depicts an example trajectory in the  $(\langle q \rangle,\langle p \rangle)$ plane, together with snapshots of the Gaussian density profile  (whose widths were rescaled for visibility). 
}
\end{figure}

As seen in Eq~(\ref{eq:cond1}), the constraint (\ref{constraint_1}) fixes
the correlation $\big<qp+pq\big> = 0$ and leads to the connection between the two work agents, Eq~(\ref{eq:mu-basic}). 
Remarkably, this coincides with a \textit{classical equipartition of energy} ($\langle p \partial_p H \rangle = \langle q \partial_q H \rangle$) \cite{Wald2015b,Wald2017} 
for our system, which has to be fulfilled dynamically at all times in the model at hand. Eq.~(\ref{constraint_2}), on the other hand, couples the evolution of the first and second moments, hence making the problem effectively non-linear (see Fig.~\ref{fig:diagram}).
We have already shown how this constraint connects to the mean-field Weiss theory 
for $h=0$. The  case $h\neq0$ can be understood as a periodic perturbation where the 'magnetic field'
$B$ is not only generated by the system $B_s = \langle q \rangle$ but where also an external
field is present $B_{ext} = h \sin \omega t$.

\subsection{Constrained dynamics}

To get a better grasp of the dynamics of the constrained system, we note that since $\langle q^2 \rangle$ and $Z$ are constants, we actually have a reduced number of variables. Since the dynamics is also unitary, a further simplification is possible, exploiting 
the fact that the purity $\mathcal{P} = \tr(\rho^2)$ of the initial state must be a conserved quantity. Hence, the evolution of the remaining degrees of freedom [Eqs.~(\ref{eq_q}), (\ref{eq_p}) and (\ref{eq_p2})] can be viewed as taking place over \textit{surfaces of constant purity}, which can  be used to  eliminate another one of the equations.

From Eqs.~(\ref{constraint_2}) and (\ref{eq:mu-basic}) we have
\begin{IEEEeqnarray}{rCl}
\label{eq:B-explicit}
B_t &=& \langle q \rangle + h \sin \omega t, 		\\[0.2cm]
\label{eq:mu-t}
\mu_t &=& \frac{\langle p^2\rangle}{m} + \langle q \rangle^2  + h \langle q \rangle \sin \omega t,
\end{IEEEeqnarray}

The dynamics is Gaussian preserving, so if the initial state is Gaussian, it will remain so for all times. 
In this case the purity can  be directly related to the first and second moments as (see appendix \ref{ap:purity}):
\begin{equation}\label{purity}
\mathcal{P} =  \frac{1}{2}\left(\langle p^2 \rangle - \langle p \rangle^2 - \langle p^2 \rangle \langle q \rangle^2\right)^{-\frac{1}{2}}.
\end{equation}
This relation can be used to express $\langle p^2 \rangle$  in terms of $\langle q \rangle$, $\langle p \rangle$ and $\mathcal{P}$. 
Moreover, as far as initial conditions are concerned, since we must always have $\langle q^2 \rangle_0 = 1$ and $Z_0 = 0$, all we need to specify are $\langle q \rangle_0$, $\langle p \rangle_0$ and $\mathcal{P}$ (which implicitly determines $\langle p^2 \rangle_0$). 

Combining Eq.~(\ref{purity}) with Eq.~(\ref{eq:mu-t})  allows us to obtain an explicit formula for the protocol for $\mu_t$: 
\begin{equation}\label{mu_t}
\mu_t =  \frac{1}{4 m\mathcal{P}^2} \bigg( \frac{1 + 4 \mathcal{P}^2 \langle p \rangle^2}{1 - \langle q \rangle^2}\bigg) + \langle q \rangle^2  + h \langle q \rangle \sin \omega t.
\end{equation}
Finally, inserting  Eqs.~(\ref{constraint_2}) and (\ref{mu_t}) into Eq.~(\ref{eq_p}), we arrive at
\begin{equation}\label{eq_p_gen}
\frac{\ud \langle p \rangle}{\ud t} = (\langle q \rangle + h \sin \omega t) (1- \langle q \rangle^2) - \frac{\langle q \rangle}{4 m \mathcal{P}^2} \bigg(\frac{ 1+ 4 \mathcal{P}^2 \langle p \rangle^2}{1 - \langle q \rangle^2}\bigg),
\end{equation}
which, together with Eq.~(\ref{eq_q}), forms a closed and highly non-linear system of equations for $\langle q \rangle$ and $\langle p \rangle$. 
Thus, even though the underlying dynamics is linear and Gaussian preserving, the implementation of the constraint leads to an effective non-linear evolution for the average position and momentum. 

\section{Equilibrium dynamics ($h = 0$)}
We begin our analysis with the simpler case $h=0$. \textcolor{black}{Since the initial conditions fulfill the 
constraints and there is neither a quench nor driving involved, this case can be viewed as 
a type of unitary equilibrium dynamics.}
\textcolor{black}{This case coincides exactly with the Weiss mean-field scenario of the quantum spherical model. 
The dissipative dynamics using Lindblad master equations has been studied in \cite{Wald2015b}. Here,
we rather shall see this as a first step to show that the constraints can lead to non-linear
dynamical effects and to lay the ground for the case $h\neq 0$.}
For $h=0$, the quantity
\begin{equation}
\Omega = \frac{\langle p^2\rangle}{m} - \langle q\rangle^2.
\label{eq:omega-simple}
\end{equation}
is also conserved, besides the purity $\mathcal{P}$. This can be seen from Eqs.~(\ref{eq_p2}) and~(\ref{eq_q}), together with $Z = 0$. Henceforth, when no confusion arises, we shall simplify the notation and write  $\langle q \rangle = q$ and $\langle p \rangle = p$. 
Exploiting the conserved quantities, we can characterise the orbits in the $(q,p)$ phase space as obeying 
\begin{equation}
m(\Omega + q^2)(1-q^2) - p^2 = \frac{1}{4\mathcal{P}^2}
\label{eq:traj}
\end{equation}
Hence, the orbit is completely determined by the values of $\Omega$ and $\mathcal{P}$. 
\begin{figure}[htbp!]
\centering
\includegraphics[width=0.22\textwidth]{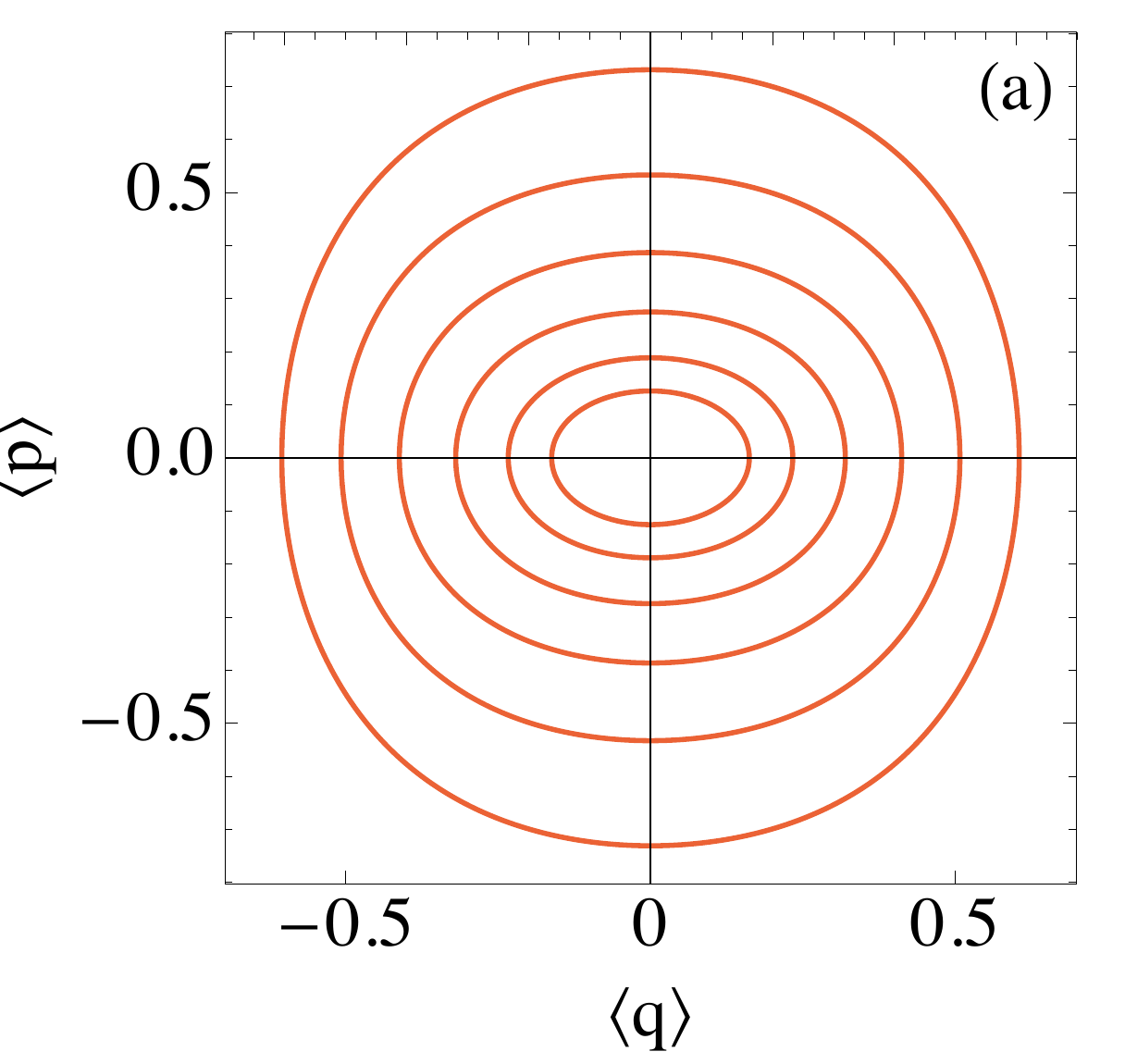}\quad
\includegraphics[width=0.22\textwidth]{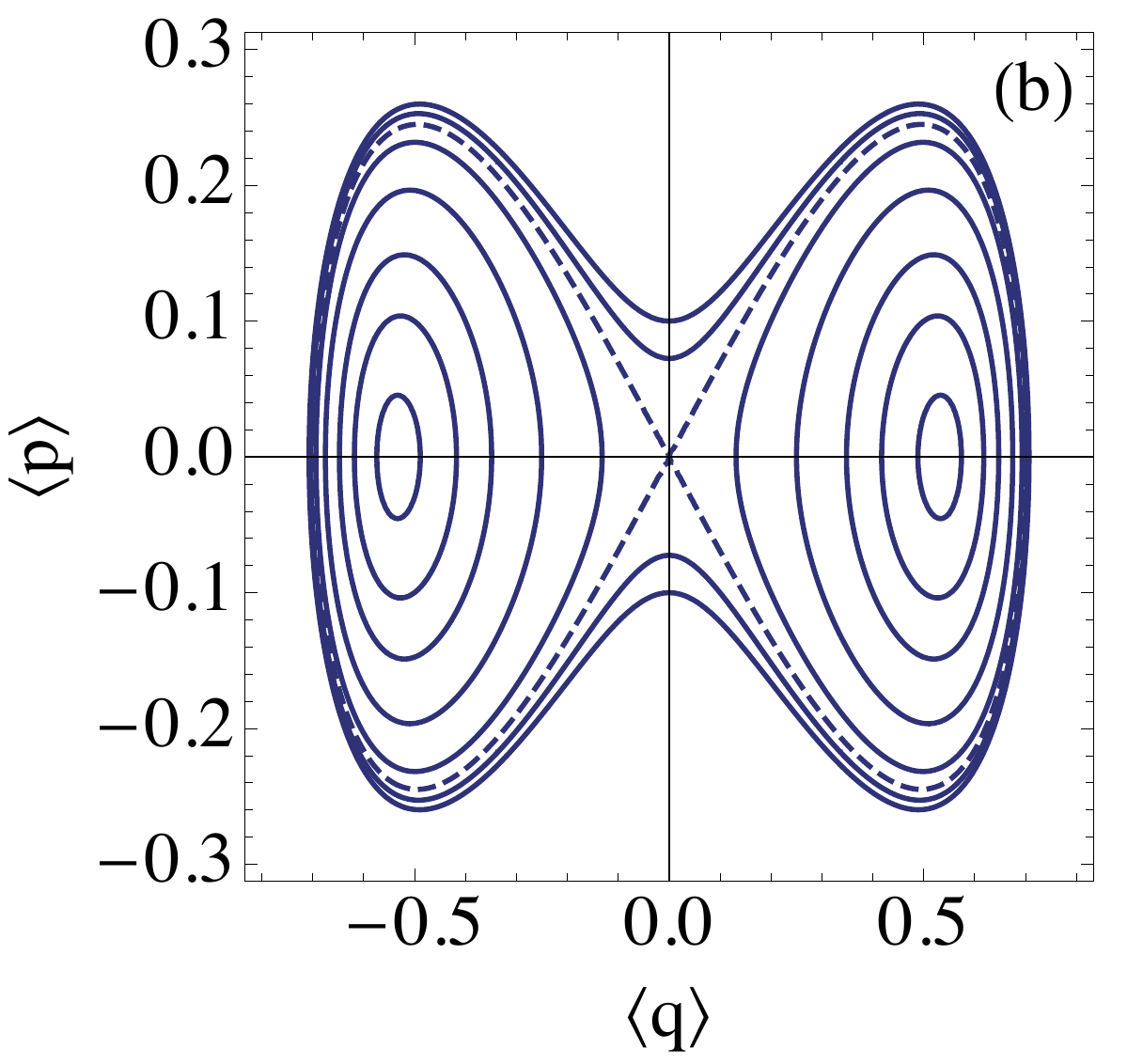}\\
\includegraphics[width=0.22\textwidth]{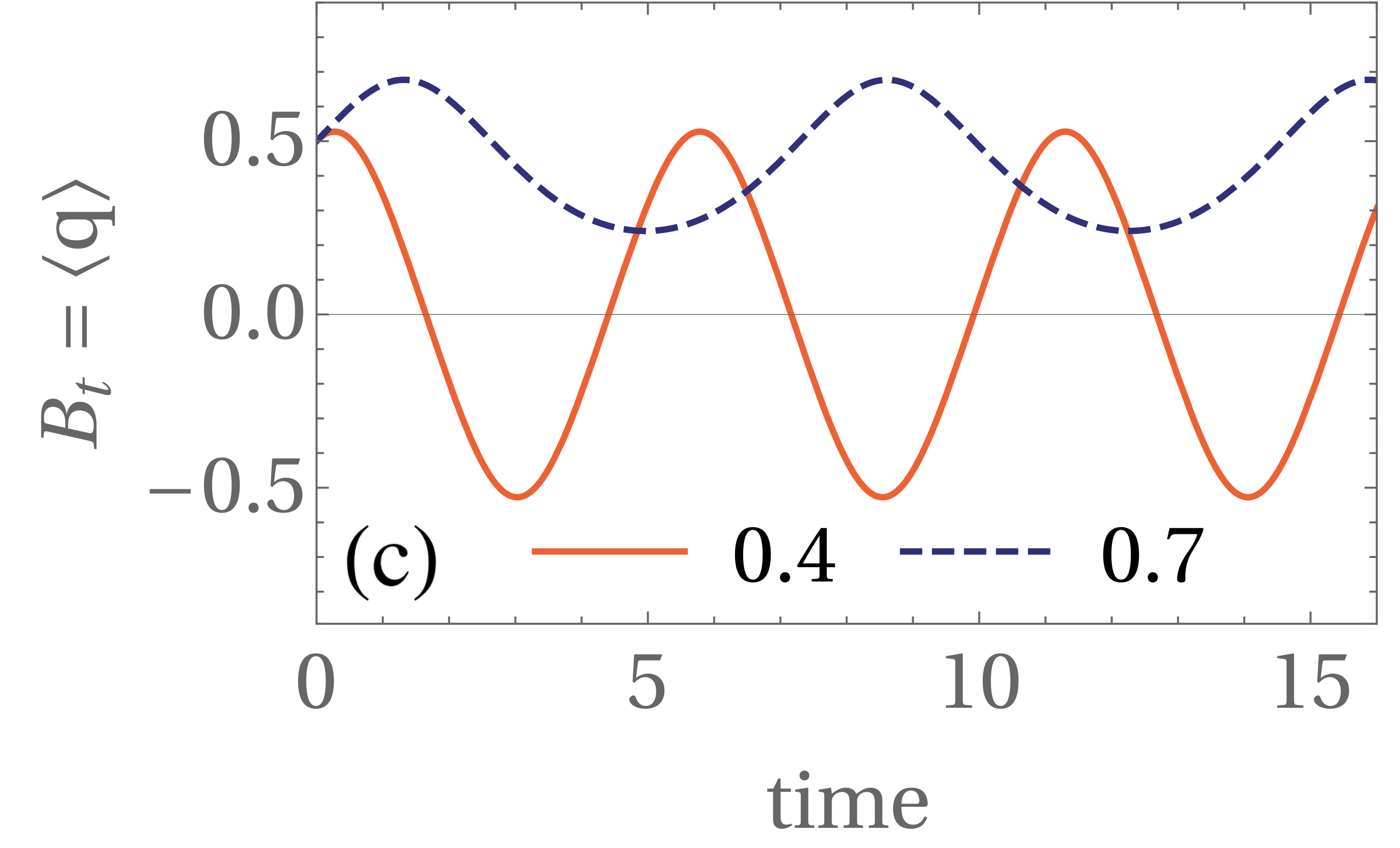}\quad
\includegraphics[width=0.22\textwidth]{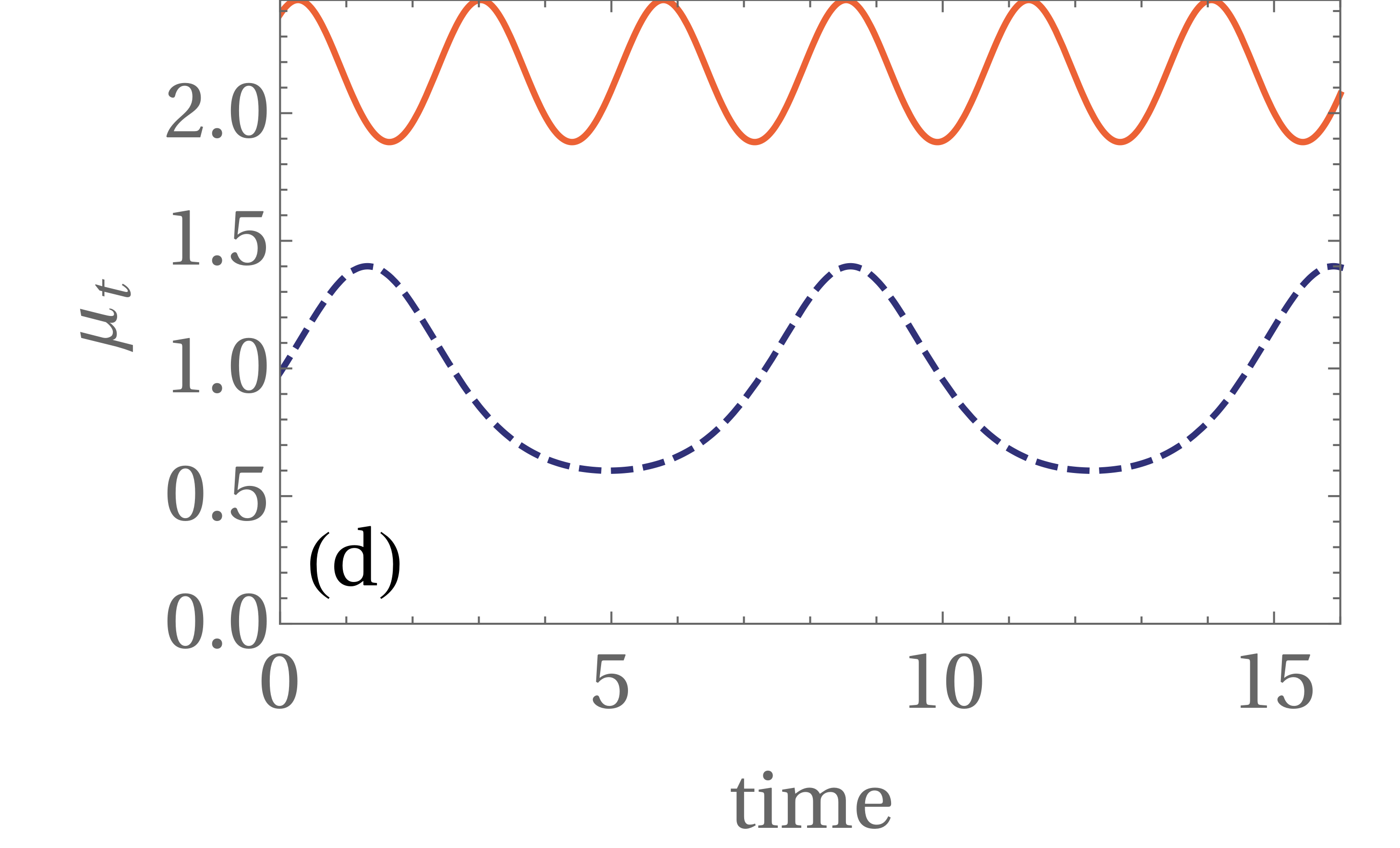}
\caption{\label{fig:trajs}
Evolution of the quantum harmonic oscillator subject to the constraint~(\ref{constraint_1}) and protocol $B_t = \langle q \rangle$, for $\omega = 1$.
(a) Orbits in the $(\langle q \rangle, \langle p \rangle)$ plane for $m = 1$ and $\mathcal{P} = 0.4$. 
(b) Same but for $\mathcal{P} = 0.7$. The homoclinic orbit crossing the origin is shown in dashed lines. 
(c-d) The protocols $B_t$ (c) and $\mu_t$ (d) for some illustrative choices of orbits with $\mathcal{P} = 0.4$ and 0.7. 
Initial conditions: $\langle q \rangle_0 = 0.5$ and $\langle p \rangle_0 = 0.2$. 
}
\end{figure}

\begin{figure*}[t!]
\centering
\includegraphics[width=0.33\textwidth]{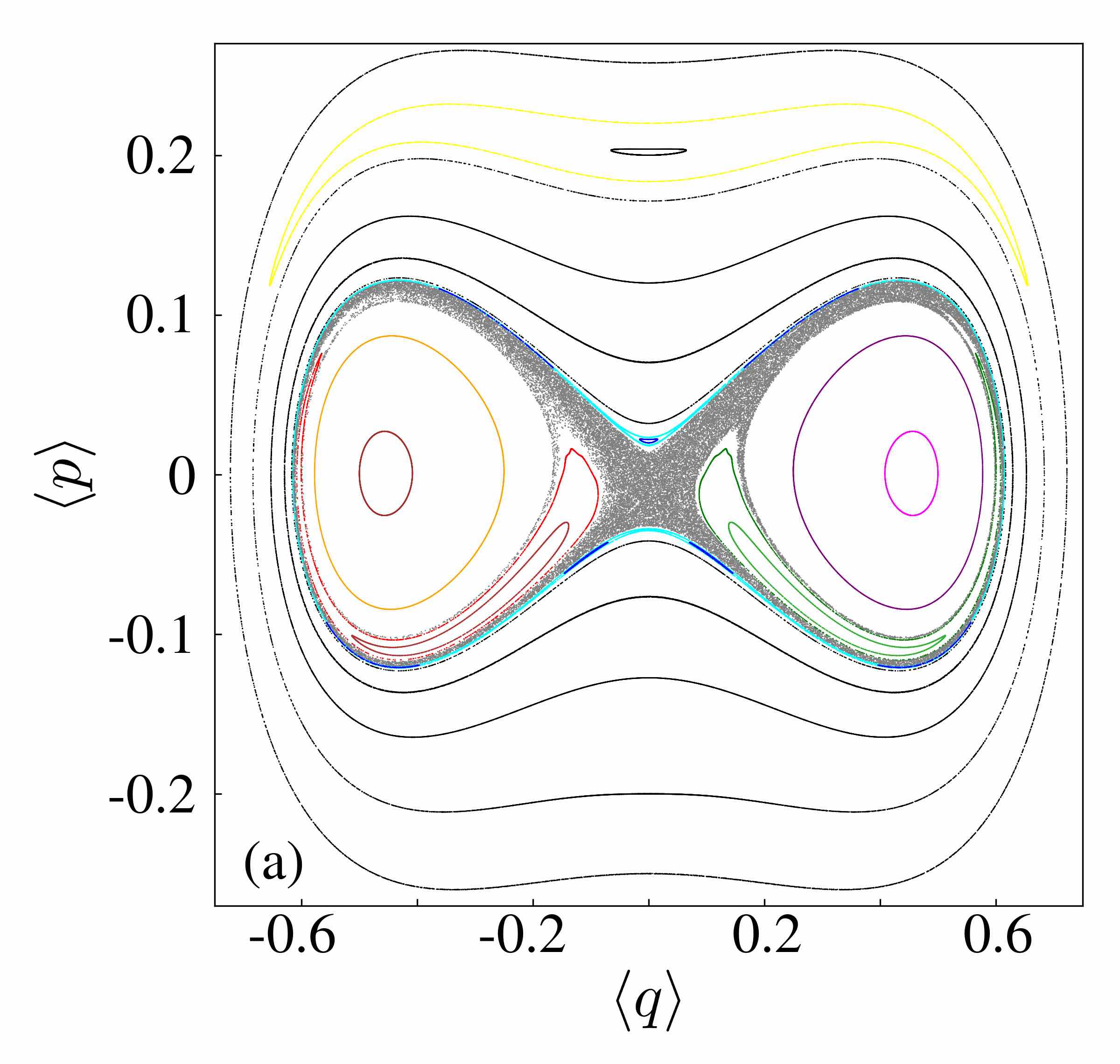}
\includegraphics[width=0.33\textwidth]{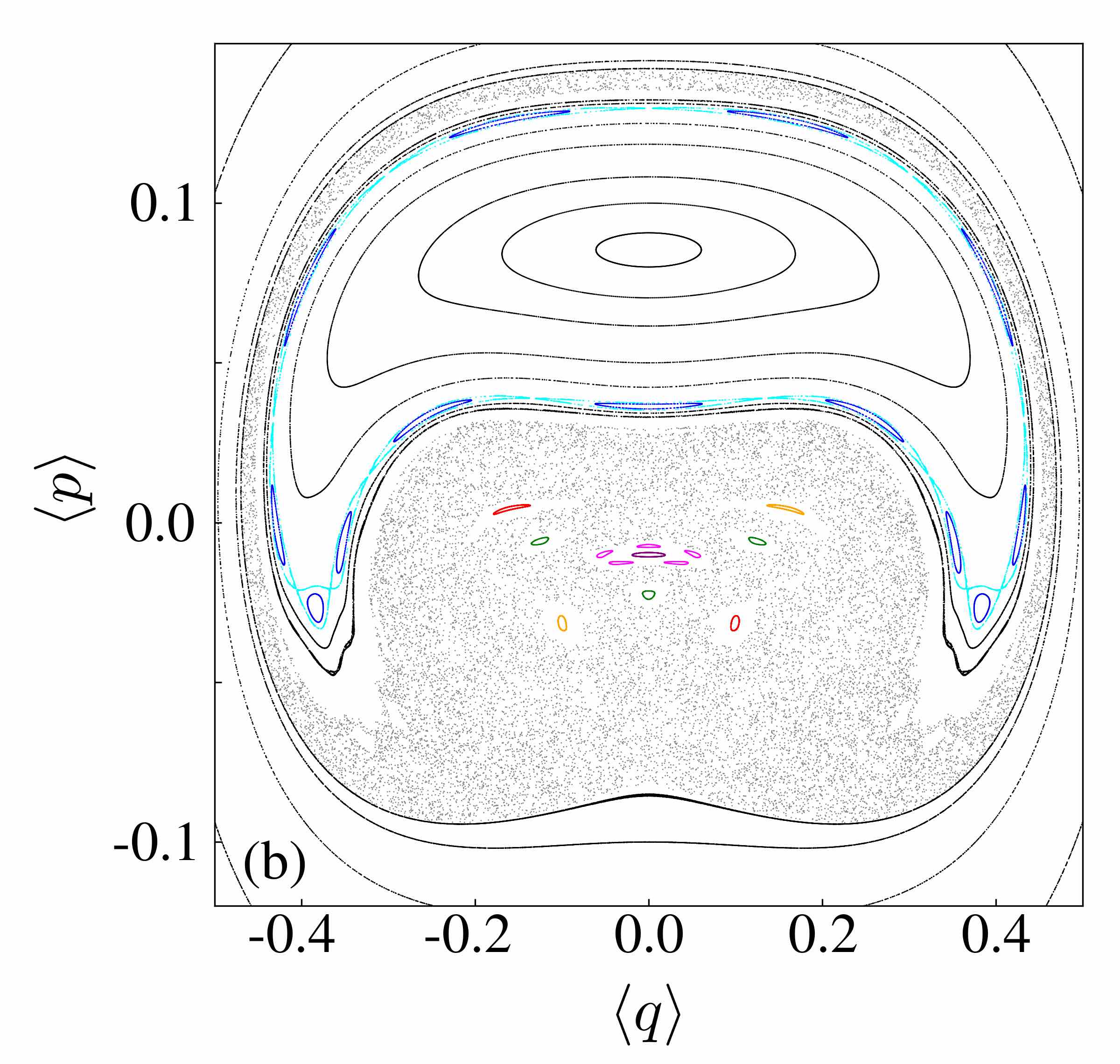}
\includegraphics[width=0.33\textwidth]{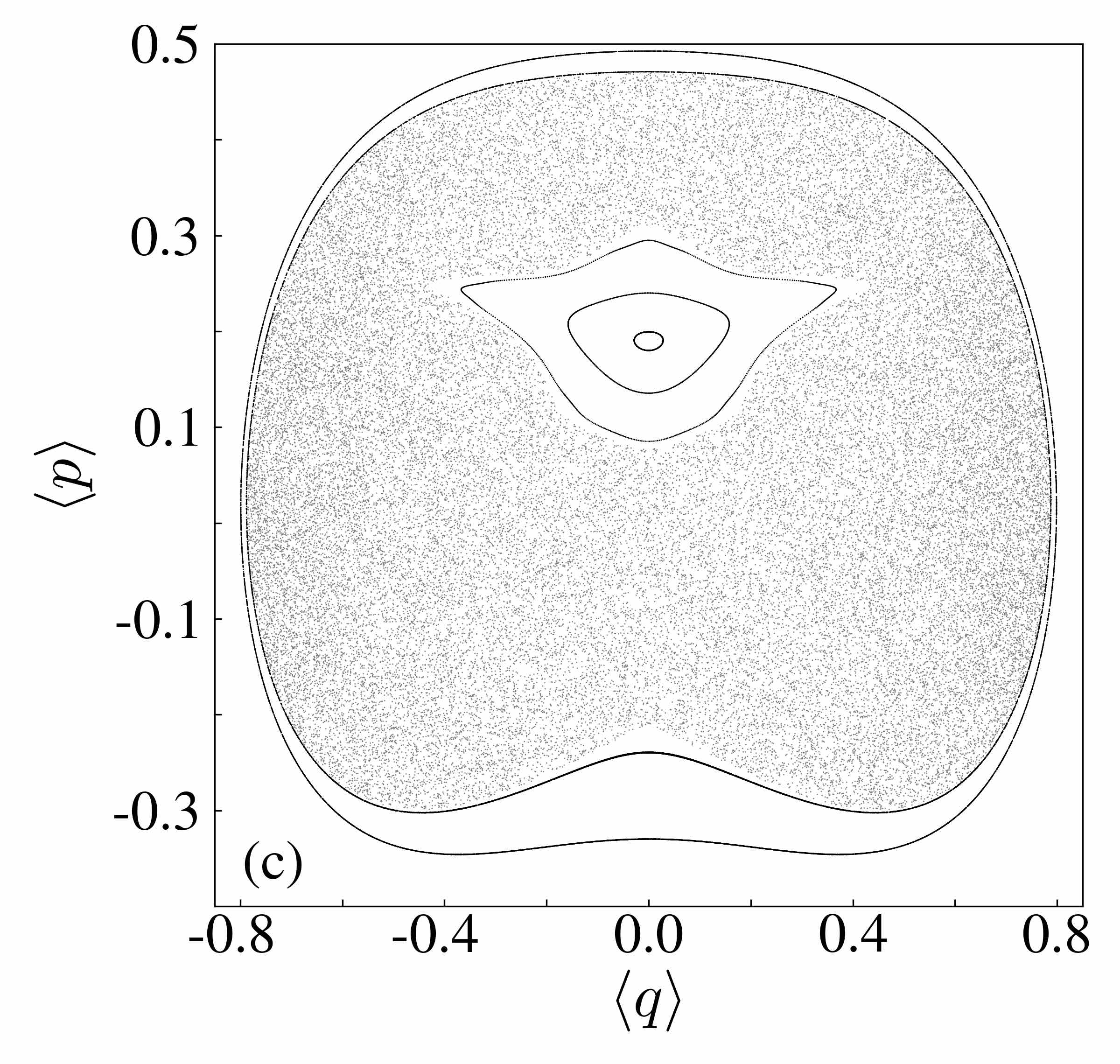}
\caption{\label{fig:chaos}
Examples of classical chaos induced by a time-dependent constraint~(\ref{constraint_2}) with $f(q,t) = q + h \sin(\omega t)$.
The plots show Poincar\'e sections in the  $(\langle q \rangle,\langle p \rangle)$ plane computed at times that are integer multiples of $2\pi/\omega$. 
Different colors correspond to different initial conditions. 
The curves were constructed with fixed $\omega = 1$, $\mathcal{P} = 1$ and different choices of $m$ and $h$:
(a) 0.4 and $10^{-3}$, (b) 0.25 and $10^{-2}$, (c) 0.4 and $10^{-1}$. 
Additional examples are shown in appendix~\ref{app:chaos}.
}
\end{figure*}

A numerical analysis of this dynamics is shown in Fig.~\ref{fig:trajs}, where we present orbits in the $(q, p)$ plane for fixed mass $m = 1$, $\omega = 1$ and different purities $\mathcal{P}$. 
In Fig.~\ref{fig:trajs}(a), where $\mathcal{P}=0.4$, only symmetric orbits covering both sides of the phase space are observed. 
Conversely, for $\mathcal{P} = 0.7$ (Fig.~\ref{fig:trajs}(b)), we see the appearance of a homoclinic solution touching the origin and acting as a separatrix between symmetric and asymmetric orbits. Finally, for the purpose of illustration, we present in Figs.~\ref{fig:trajs}(c-d) examples of the protocols $B_t$ and $\mu_t$, which must be implemented in the actual evolution in order to enforce the constraint.

The homoclinic solution is found when the curve going through $(0,0)$ crosses $p=0$ in 2 other points. 
From Eq.~(\ref{eq:traj}), this means that homoclinic solutions exist for $\Omega = \nicefrac{1}{4m\mathcal{P}^2} < 1$. From Eq.~(\ref{eq:omega-simple}) we have that $\Omega>-1$, so the existence of asymmetric solutions depends only on the relationship between the purity and the mass. More precisely, asymmetric solutions exist iff
%
%
\begin{equation}\label{eq:dqpop}
\mathcal{P} < \mathcal{P}_c = \frac{1}{2\sqrt{m}}
\end{equation}
In terms of many-body quantum systems, Eq.~(\ref{eq:dqpop}) presents a \textit{dynamical 
quantum phase transition at mean field level}, in the sense that a dynamical order parameter
\begin{equation}
 \bar{q} = \lim_{T\to\infty}T^{-1} \int_0^T \left< q\right>(\tau) \ d\tau
\end{equation}
can be defined that distinguishes between a symmetric and a symmetry-broken phase. Such dynamical
order parameter transitions are somewhat less studied than the more common Loschmidt echo dynamical 
quantum phase transitions \cite{Heyl_2019} although they are closely related in certain 
cases \cite{PhysRevLett.120.130601}.

The transition in Eq~(\ref{eq:dqpop}) explicitly relies on the 
introduction of two distinct constraints and enriches the behaviour of this model as it shows not
only this dynamical quantum phase transition but as well a steady state dissipative transition
\cite{Wald2015b}. This remarkably rich dynamical behaviour may lead to a variety of possibilities
for further explorations such as interplay between dynamical and dissipative transitions which can be explicitly studied in this simple toy model. We hope to return to this point in a future work.
\section{Periodically driven dynamics and robustness of the approach}

Next we consider the effects of adding a sinusoidal forcing on top of $B_t$, that is, we set $h \neq 0$ in Eq.~(\ref{eq:B-explicit}). \textcolor{black}{As we already mentioned, this study can be viewed 
as an exploration of many-body quantum Floquet dynamics in the mean-field regime. From this 
viewpoint, the mass of the oscillator plays the role of quantum fluctuations in the system, in the 
sense that $m\to \infty$ corresponds to the classical spherical model. Conversely, the purity
describes the mixing of the initial quantum state. Roughly speaking, we thus find two sources of
quantumness in the system and without further justifaction, we choose to study the dynamical 
behaviour for constant purity $\mathcal{P} = 1$ and different strenghts of quantum fluctuations $1/m$ around the critical value $m_c = \nicefrac{1}{4}$.}

One of the main results we wish to emphasize from this analysis is that 
\textit{time-dependent constraints can lead to classical chaotic behavior} for the first moments $q$ and $p$. 
This is illustrated in Fig.~\ref{fig:chaos} where we show Poincar\'e sections of the $(q,p)$ plane for different choices of parameters and initial conditions. 
As can be seen, imposing the constraint leads to a remarkably rich set of responses of the system.


We observe three distinct dynamical phases for the periodically driven 
mean-field model
\begin{enumerate}
 \item \emph{a symmetric (or paramagnetic) phase}, where orbits are quasi-periodic with $\bar{q} = 0$,
 \item \emph{a broken symmetry (or ferromagnetic) phase}, where orbits are quasi-periodic with $\bar{q} \neq 0$,
 \item \emph{a chaotic phase}, where $\bar{q} = 0$ and orbits are completely aperiodic.
\end{enumerate}

\begin{figure}[!h]
 \includegraphics[width=.23\textwidth]{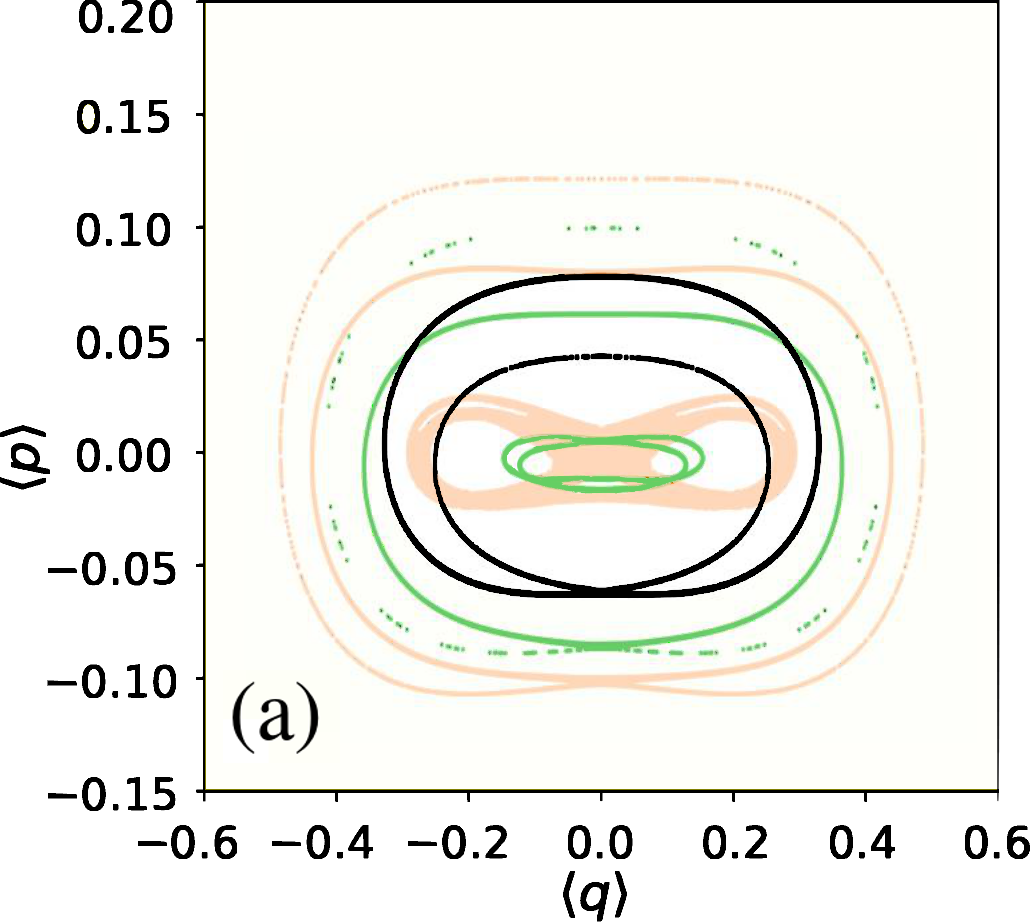}
  \includegraphics[width=.23\textwidth]{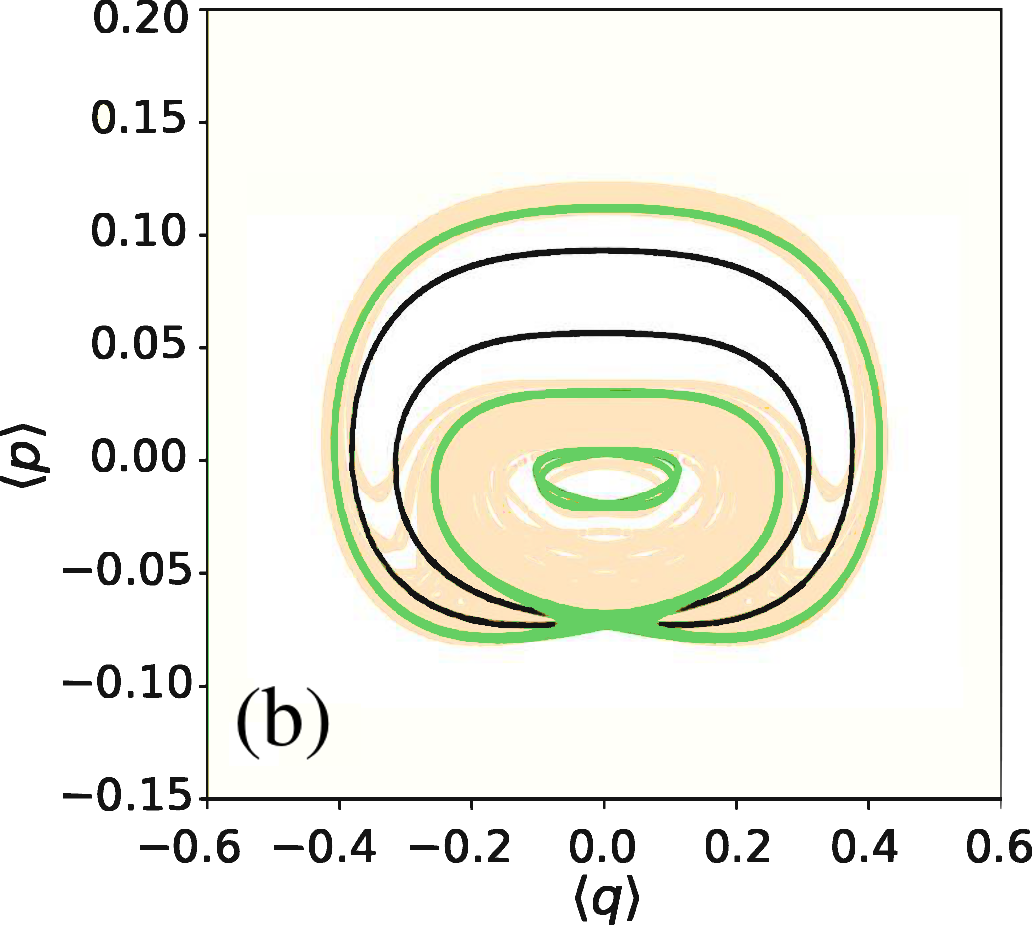}
 \caption{\label{fig:phasediagram} Chaotic region in phase space:
 \emph{left panel}: for $h=0.001$ and different masses
 $m=0.23$ (black), 0.25 (green / dark gray), 0.27 (beige / light gray); \emph{right panel}: for $m=0.24$ and different driving strengths
 $h=0.001$ (black), 0.005 (green / dark gray), 0.008 (beige / light gray)}
\end{figure}

While the symmetric as well as the broken symmetry phase were expected since they are already present in 
the scenario $h=0$, the chaotic phase that is induced by the periodic driving is indeed surprising from 
the point of view of a many-body system.
In figure~\ref{fig:phasediagram} we depict phase diagrams that distinguish between chaotic and non-
chaotic regions. We see that the chaotic regions are indeed extensive and one can thus expect to 
observe these in a Floquet-type setup since it is not a fine-tuned effect.

To understand this chaotic behavior, first note that the trajectories in the case $h=0$ (defined by Eq.~(\ref{eq:traj})) are closed, meaning the system is periodic. This means that the system in the absence of a drive is a nonlinear oscillator. Once we add the drive, the situation becomes similar to other periodically driven nonlinear oscillators that display chaotic behaviour for certain ranges of parameters, like the Duffing and the van der Pol oscillators \cite{duffing-1, duffing-2, van-der-pol}. This explains the behaviour for moderate and large values of $h$, but another feature that contributes for the presence of chaos even for very low $h$ is the existence of the homoclinic trajectory for $h=0$ (shown in Fig.~\ref{fig:trajs}(b)). After we add the drive, the periodic solutions in Fig.~\ref{fig:trajs} will have associated quasi-periodic solutions that will correspond to invariant tori in an extended phase space, where we added time as an extra direction with periodic boundary conditions. The problem is that the homoclinic solution would be in the intersection of 2 of these tori, which is a forbidden structure (for the same reason different trajectories cannot cross in a phase space). So what happens instead is that these 2 tori ``merge'' into a single aperiodic (hence chaotic) solution and as $h$ grows, this aperiodic solution engulfs a larger region.

Notwithstanding, the quantum mechanical evolution continues to be linear and Gaussian preserving. This happens because of the way our constraints are implemented, through $\mu_t$ and $B_t$ which are themselves chaotic. This means that we can think of the whole system as an ordinary bosonic mode subject to an external agent that behaves chaoticaly, but ensures $\langle q^2 \rangle = 1$. Given this unusual state of affairs and the usual sensitivity of chaotic dynamics to initial conditions and perturbations, a natural question is whether such an implementation would be feasible in practice.
We next show, by means of a numerical analysis, that the answer to this question is positive.

\begin{figure}[ht!]
\centering
\includegraphics[width=0.23\textwidth]{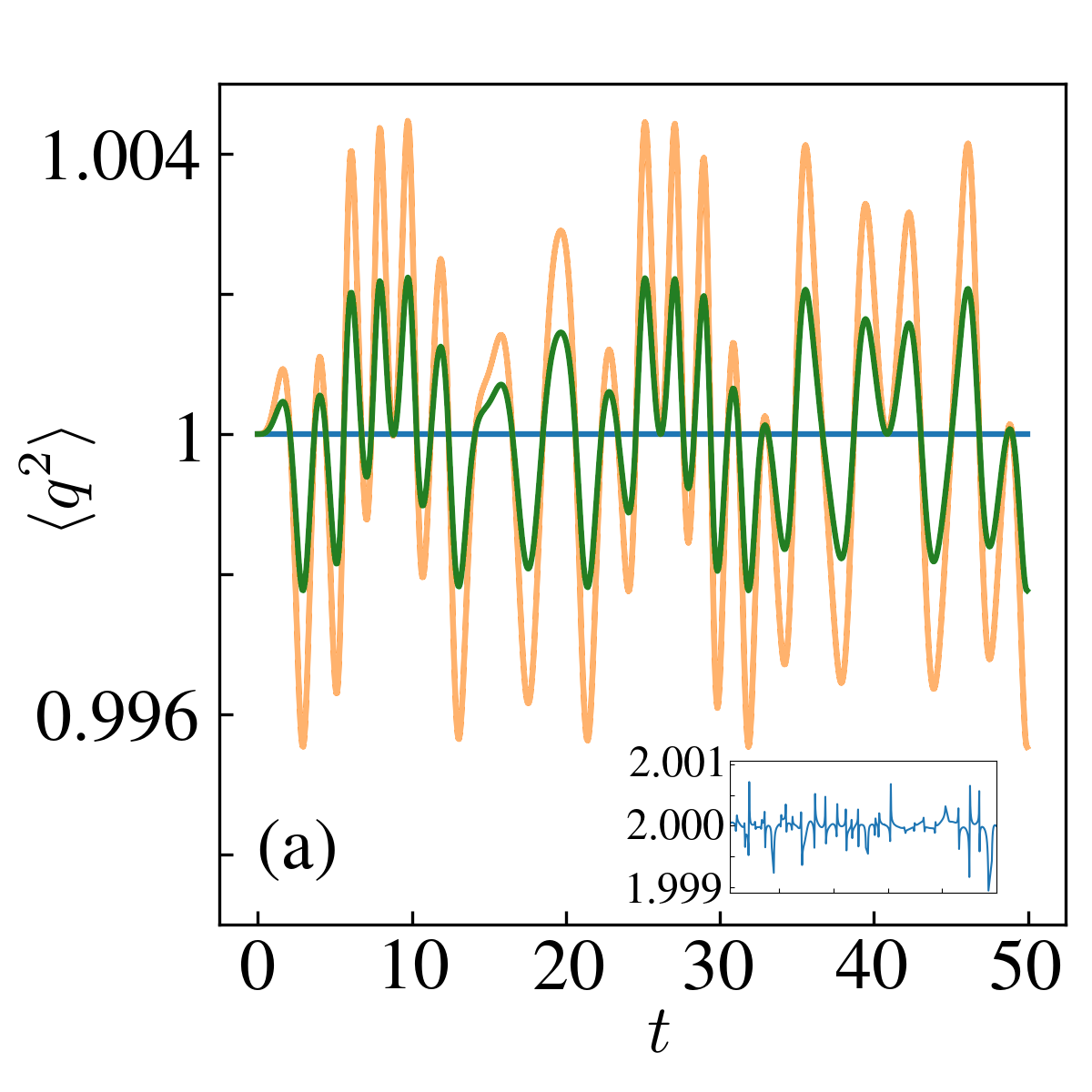} \quad
\includegraphics[width=0.23\textwidth]{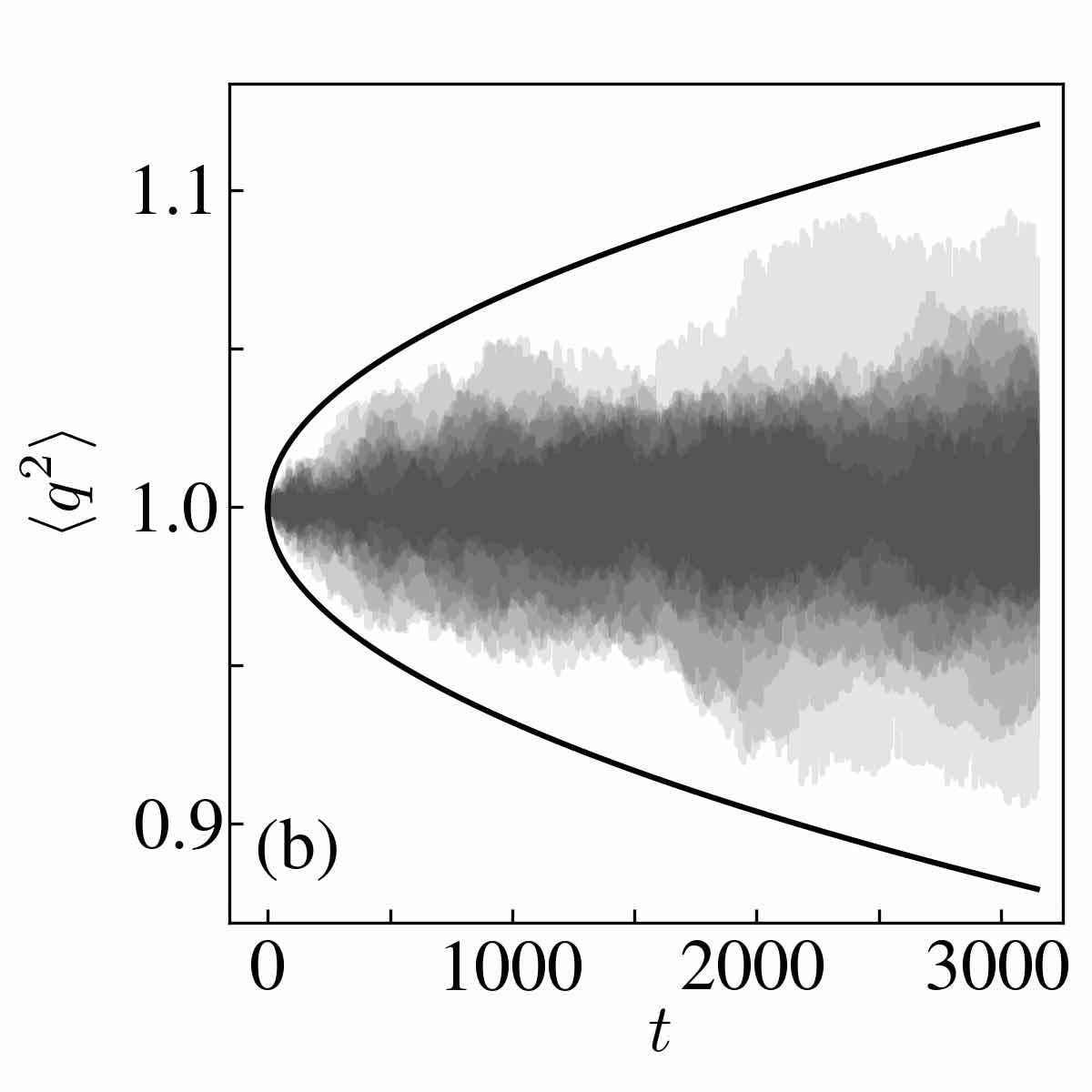}
\caption{\label{fig:robust}
Robustness of the protocol implementation to potential sources of error in the case of the time-dependent constraint used in Fig.~\ref{fig:chaos}.
In both figures we present curves of $\langle q^2 \rangle$ as a function of time which, in an ideal case, should equal  unity due to the constraint in Eq.~(\ref{constraint_1}).
(a) Error due to a protocol designed for the wrong initial conditions: $\langle q\rangle_0^{\text{act}}- \langle q \rangle_0 = \langle p\rangle_0^{\text{act}} - \langle p \rangle_0 = 10^{-3}$ and $2 \times 10^{-3}$. 
In this case the total error does not scale with time and is also linearly proportional, at all times, to the original error. 
The inset shows the ratio of the two errors, which remains close to 2 at all times. 
(b) Error due to the presence of a random noise in the protocol $\mu_t$.
Details on how this noise is implemented are give in appendix~\ref{ap:robustness}.
The curve shows $\langle q^2\rangle$ for several trajectories. 
In this case the error scales at most as $\mathcal{O}(t^{1/2})$, illustrated by the black-solid line. 
Other parameters were $m = 0.4$, $h = 0.1$, $\omega = 1$ and $\mathcal{P} = 1$. 
}
\end{figure}

There are two main potential sources of error in the implementation of a protocol. 
The first is an error in the initial conditions. 
That is, one may design a protocol meant for a given  $\langle q \rangle_0$, which does not coincide exactly with the actual initial condition $\langle q\rangle_0^{\text{act}}$ (and similarly for the other initial conditions). 
However, the fact that the chaos in our system is being imposed on top of a linear evolution means that the overall error will simply be  proportional to the initial error $\langle q\rangle_0^{\text{act}} -  \langle q \rangle_0$ and, most importantly, 
will not increase exponentially with time.
This is illustrated in Fig.~\ref{fig:robust}(a) where we show the time evolution of $\langle q^2 \rangle $ (which ideally should equal unity according to Eq.~(\ref{constraint_1})) assuming different errors in $\langle q\rangle_0^{\text{act}} -  \langle q \rangle_0$. 
As can be seen, doubling the initial error simply doubles the error at a time $t$ (which is also emphasized in the inset of Fig.~\ref{fig:robust}(a)).

Another potential source of error are imprecisions in the protocol itself. 
This question is less trivial to address, as it relates to the concept of structural stability of a dynamical system. 
We simulate this effect by introducing a random noise of varying intensity in the protocol $\mu_t$. 
Details on how this is implemented are given in appendix \ref{ap:robustness} and a numerical illustration is shown in Fig.~\ref{fig:robust}(b). 
As can be seen, a numerical error in the protocol $\mu_t$ leads to an accumulation of the error which scales, at most, with order $\mathcal{O}(t^{1/2})$, illustrated by the black lines (see appendix \ref{ap:robustness} for a more in depth analysis). 
Thus, even though the error does accumulate in this case, it is sub-linear in time and not exponential, so it may still be manageable provided the experimental running times are not too long.

\section{Discussion}
Imposing external constraints on the evolution of quantum systems is a decades-old idea, motivated by both practical aspects in quantum control and fundamental aspects, such as quantum gravity \cite{Giampaolo2018} or its important links to many-body quantum physics.
However, due to the inherent difficulty related to the non-commutativity of quantum mechanical operators, it has never enjoyed the breadth and scope of its classical counterpart. 
Instead, most of the advances in this direction have actually taken place indirectly in the field of quantum control, in particular with techniques such as shortcuts to adiabaticity and dynamical decoupling.
The main goal of this paper was to show that these techniques can potentially be extended to formulate a consistent theory of  constrained  quantum dynamics. 
Our focus has been on the case of a single quantum harmonic oscillator, due both to its simplicity and to its natural appeal in several experimental platforms, such as trapped ions and optomechanics. 
We formulated our system in terms of Eqs (\ref{H}), (\ref{constraint_1}) and
(\ref{constraint_2}) and sketched how such a rather simple constrained quantum evolution can 
be viewed as a mean-field study of high-dimensional many-body quantum dynamics.
In the case of strictly time-independent constraints, we observed a dynamical quantum phase 
transition in the unitary regime that is induced by the interaction of \textit{both} constraints. We then
turned to study the effects that periodic driving has on such a transition. We find that for large
portions of the phase space, there is a new phase emerging in which orbits are not periodic but rather 
follow classical chaotic motion. This fact paired with the stability against the two experimentally
most relevant errors yields that the chaotic phase we are showcasing here can indeed be observed in 
an experimental setup and does not rely on fine-tuning.

A similar approach can of course also be developed in the case of dichotomic systems.
For instance, in  Ref.~\cite{Gustavsson2009} the authors studied the dynamics of two qubits under the constraint that they remain disentangled throughout. 
Similar extensions are also possible for continuous variables. 
Indeed, even cases as simple as 2 bosonic modes already open up an enormous number  of possibilities. 
For instance, with two bosonic modes one could implement a Kapitza pendulum \cite{Lerose2018a}, or investigate variations of this in which the constraints are not only among the averages, but also involve fluctuations.

{\bf Acknowledgements - }
The authors acknowledge Gabriele de Chiara and Frederico Brito for stimulating discussions. 
GTL acknowledges the financial support of the S\~ao Paulo Research Foundation, under project 2016/08721-7.
GTL acknowledges SISSA and SW acknowledges USP for the hospitality.
FS acknowledges partial support from of the Brazilian National Institute of Science and Technology of Quantum Information (INCT-IQ) and CNPq (Grant No. 307774/2014-7)


\appendix

\section{Purity of a Gaussian system}
\label{ap:purity}

We now briefly comment on the expression for the purity $\mathcal{P}$ used in the main text [Eq.~(\ref{purity})].
The covariance matrix for a single bosonic mode is defined as 
\[
\sigma = 
\begin{pmatrix}
\langle  q^2 \rangle-\langle q\rangle^2 & Z - \langle q \rangle \langle p \rangle \\[0.2cm] 
Z - \langle q \rangle \langle p \rangle & \langle  p^2 \rangle -\langle p \rangle^2
\end{pmatrix},
\]
where, recall, $Z = \frac{1}{2} \langle qp+pq\rangle$. 
In view of Eq.~(\ref{constraint_1}) of the main text, we have $\langle q^2 \rangle =1$. 
Moreover, as seen from Eq.~(\ref{eq:cond1}), we must also have $Z = 0$. 
Thus, the covariance matrix becomes 
\[
\sigma =
\begin{pmatrix}
1-\langle q \rangle^2 		&	-\langle q \rangle\langle p \rangle	\\[0.2cm]
-\langle q \rangle\langle p \rangle	&	 \langle p^2 \rangle - \langle p \rangle^2
\end{pmatrix}.
\]
For Gaussian states, it is well know that the purity may be written as 
\[
\mathcal{P} = \frac{1}{2\sqrt{|\sigma|}}.
\]
Carrying out the computation then leads to Eq.~(\ref{purity}) of the main text. 

\section{Method used for the study of robustness against protocol errors}
\label{ap:robustness}

\begin{figure*}[t]
\includegraphics[width=\textwidth]{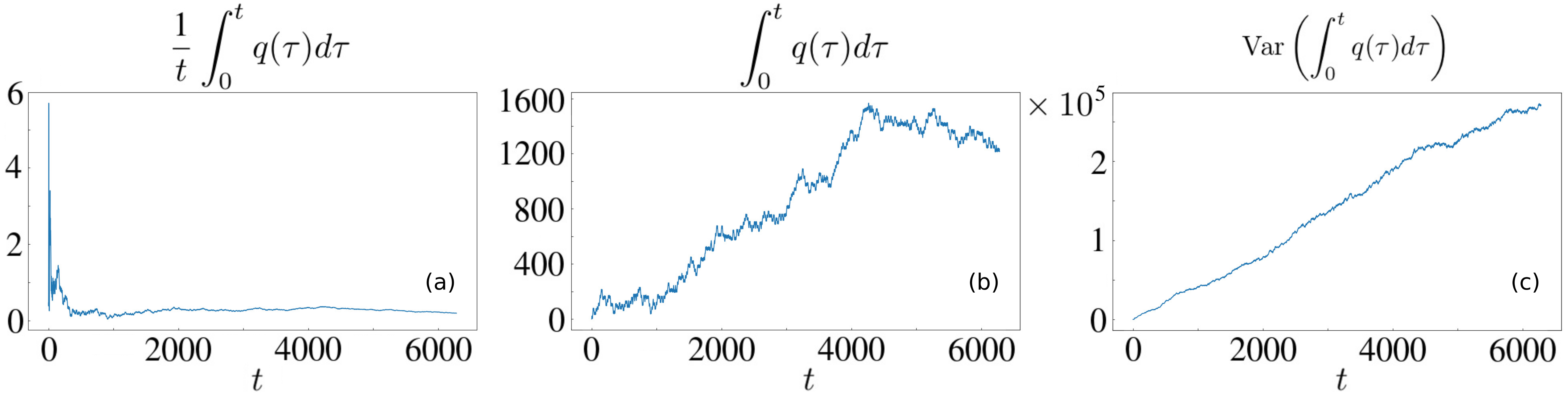}
        \caption{\label{fig-ap:q}
                    (a): Average of $\langle q \rangle$ over time for a single realization of the chaotic orbit.
                    (b): The integral over time of $\langle q \rangle$ for a single realization of the chaotic orbit, showing the resemblance to a Wiener process.
                    (c): As it happens with the Wiener process, if we consider the sample variance of several realizations of the chaotic orbit integrated over time, we get a linear dependence with time.}
        \label{fig:gull}
\end{figure*}

In this section we detail the perturbation used for the study presented in Fig. 4(b). According to our approach, given an initial condition $\langle q \rangle$ and $\langle p \rangle$ and a state with purity 1, there are protocols $\mu_t$ and $B_t$ that, if followed precisely, lead to the constraint $\langle q^2 \rangle = 1$. However in an experimental setting this might not be possible and instead we will have
\begin{align*}
\mu_{\mathrm{experiment}} &= \mu_{\mathrm{theory}} + \delta_{\mu} \\
B_{\mathrm{experiment}} &= B_{\mathrm{theory}} + \delta_{B}.
\end{align*}
where $B_{\mathrm{theory}}, \mu_{\mathrm{theory}}$ are the $B_t$ and $\mu_t$ predicted for the initial conditions used and the $\delta$ are  (presumably) small sources of error. We are interested in understanding the effect these perturbations can have in our constraint and for how long we can expect it to hold if the perturbations $\delta_{\mu}$ and $\delta_{B}$ have a size about $\varepsilon$. Unfortunately, the full problem (understanding the robustness against all possible choices of perturbations) cannot be feasibly treated, so we must choose some kind of representative noise. We want our noise to have the following properties:
\begin{itemize}
\item Be continuous.
\item Have 0 average along time.
\item Have a finite correlation time.
\item Be mostly bounded, so that $|\delta| < \varepsilon$ most of the time.
\end{itemize}
An obvious candidate would be to make $\delta$ an Ornstein-Uhlenbeck process, however this turns our problem into a stochastic differential equation, which is way more costly to solve than an ordinary equation and would make
the simulations quite time demanding.

We decided instead to follow a physically meaningful model based on a chaotic attractor as our source of noise. More precisely we set our parameters to $h = 0.1$ and $m = 0.4$. As it can be seen in the Poincar\'e section in Fig.~3(c) of the main text, trajectories starting close to the origin $\langle q \rangle = \langle p \rangle = 0$ are chaotic in this case. So for each simulation we chose random initial conditions close to the origin and used $\delta = \varepsilon \langle q \rangle$ for this chaotic orbit (meaning that we were integrating numerically simultaneously one copy of the non-linear system to obtain $B_{\mathrm{theory}}$ and $\mu_{\mathrm{theory}}$, two other copies of this system to obtain $\delta_{\mu}$ and $\delta_{B}$ and one copy of the linear system to investigate how the constraints were being affected by the perturbation).

$\varepsilon\langle q \rangle$ in the chaotic regime clearly satisfies the properties of being continuous and bounded. Furthermore, the sensitivity to initial conditions guarantees the finite correlation time. The average along time being 0 is less obvious but it is a consequence of the model having a $(\langle q \rangle, t) \rightarrow (-\langle q \rangle, -t)$ symmetry, meaning that the invariant measure of the chaotic region must be an even function of $\langle q \rangle$. This can also be checked from simulations (figure \ref{fig-ap:q}(a)).
Finally, we also comment that the time integral  of $\langle q \rangle$ in the chaotic phase displays properties akin to that of a random walk, which serves as a further indication that $\langle q \rangle$ is a good source of ``random'' noise (figures \ref{fig-ap:q}(b) and \ref{fig-ap:q}(c)).

%
%

\begin{figure}[!h]
\centering
\includegraphics[width=0.4\textwidth]{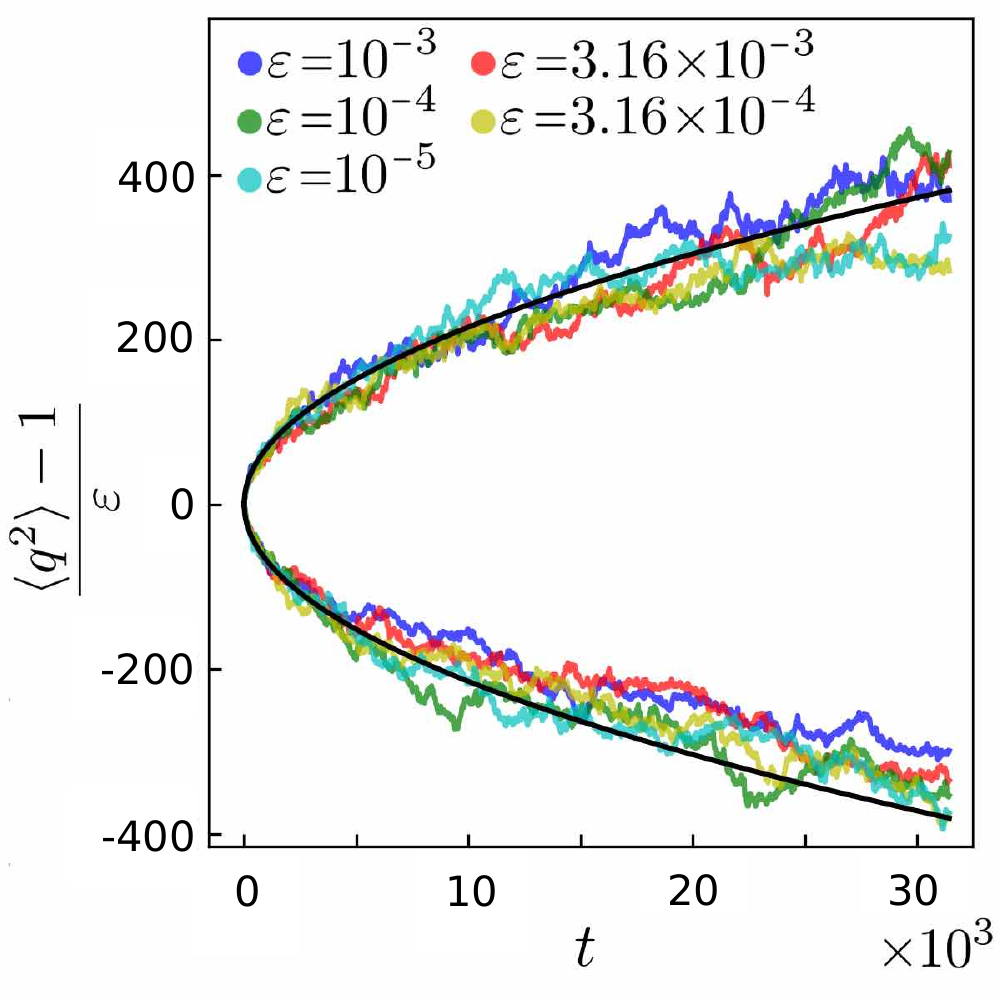}
\caption{
\label{fig:construction_black}
Construction of the black-solid curve in Fig.~\ref{fig:robust}(b) of the main text and analysis of the effect of the noise intensity $\varepsilon$. 
Each colored curve represents the maximum error found for a given value of $\epsilon$, considered over 100 stochastic trajectories. 
From these curves we find that the error $\langle q^2 \rangle -1$ scales linearly in $\epsilon$, so that all curves can be collapsed into a single plot. 
An additional fitting of these curves with a power law behavior $t^\alpha$ then reveals the exponent $t^{1/2}$ reported in the main text.
}
\end{figure}

With this choice of noise, we then reproduce the simulations several times, always using a different seed. 
Each stochastic run produces a curve of the form of Fig.~\ref{fig:robust}(b) in the main text. 
Considering, for 100 realizations, the maximum deviations of $\langle q^2\rangle$ below and above 1, we constructed the black-solid curve in Fig.~\ref{fig:robust}(b) of the main text, giving an estimate of the maximum allowed errors given a noise intensity $\epsilon$ and how this maximum error scales with time (see Fig.~\ref{fig:construction_black}).

\section{Emergence of the chaotic solution}\label{app:chaos}
\begin{figure*}[t]
\centering
\includegraphics[width=0.3\textwidth]{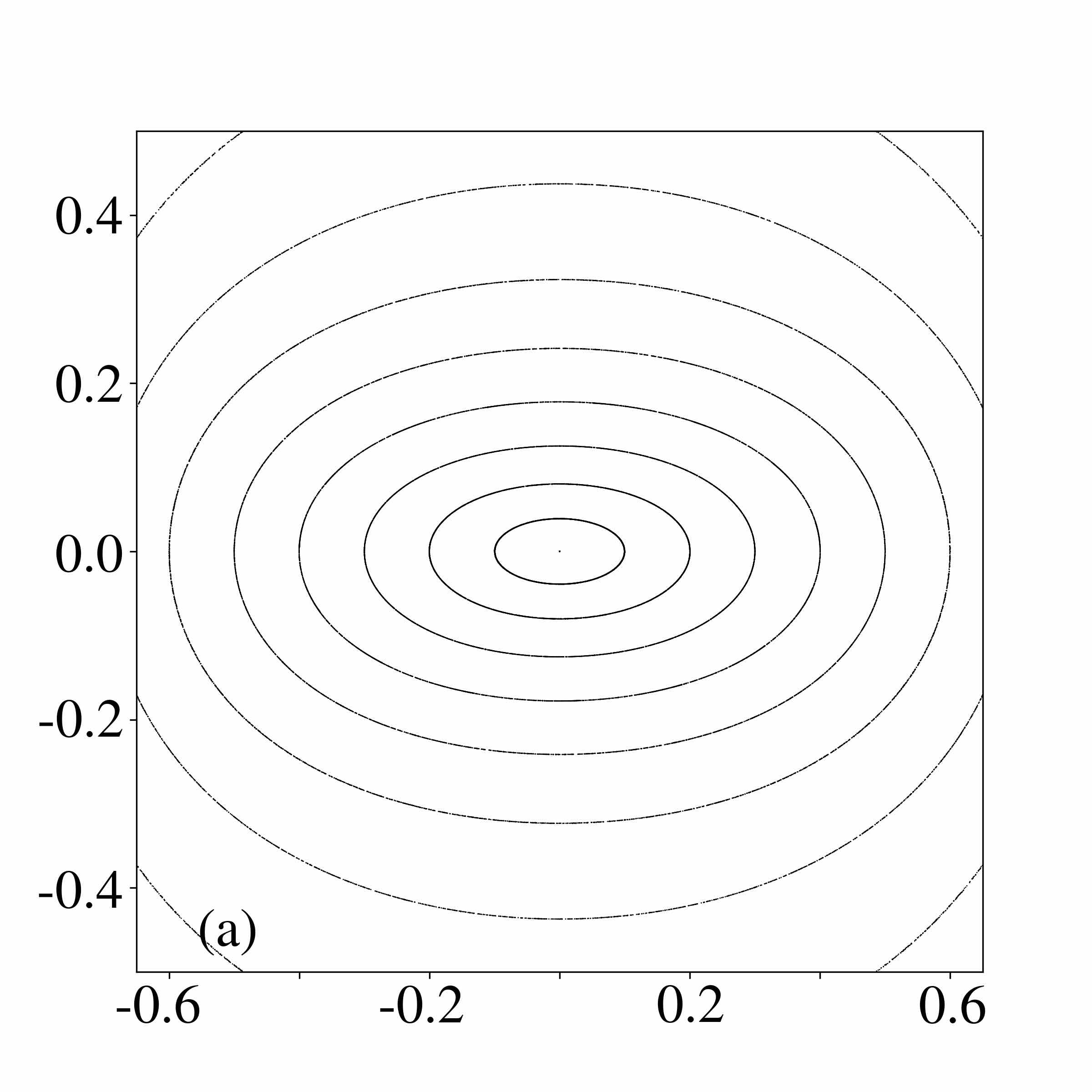}
\includegraphics[width=0.3\textwidth]{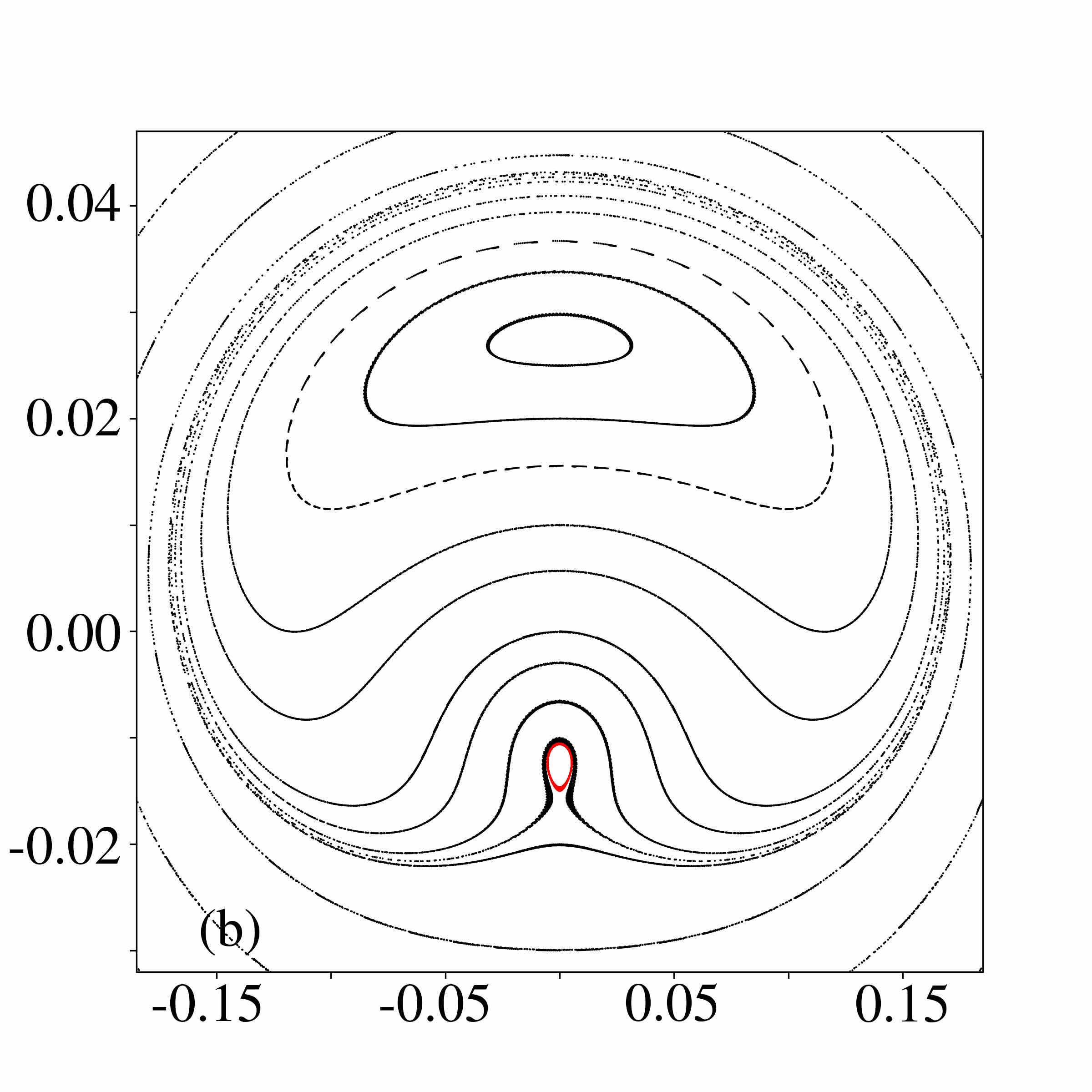}
\includegraphics[width=0.3\textwidth]{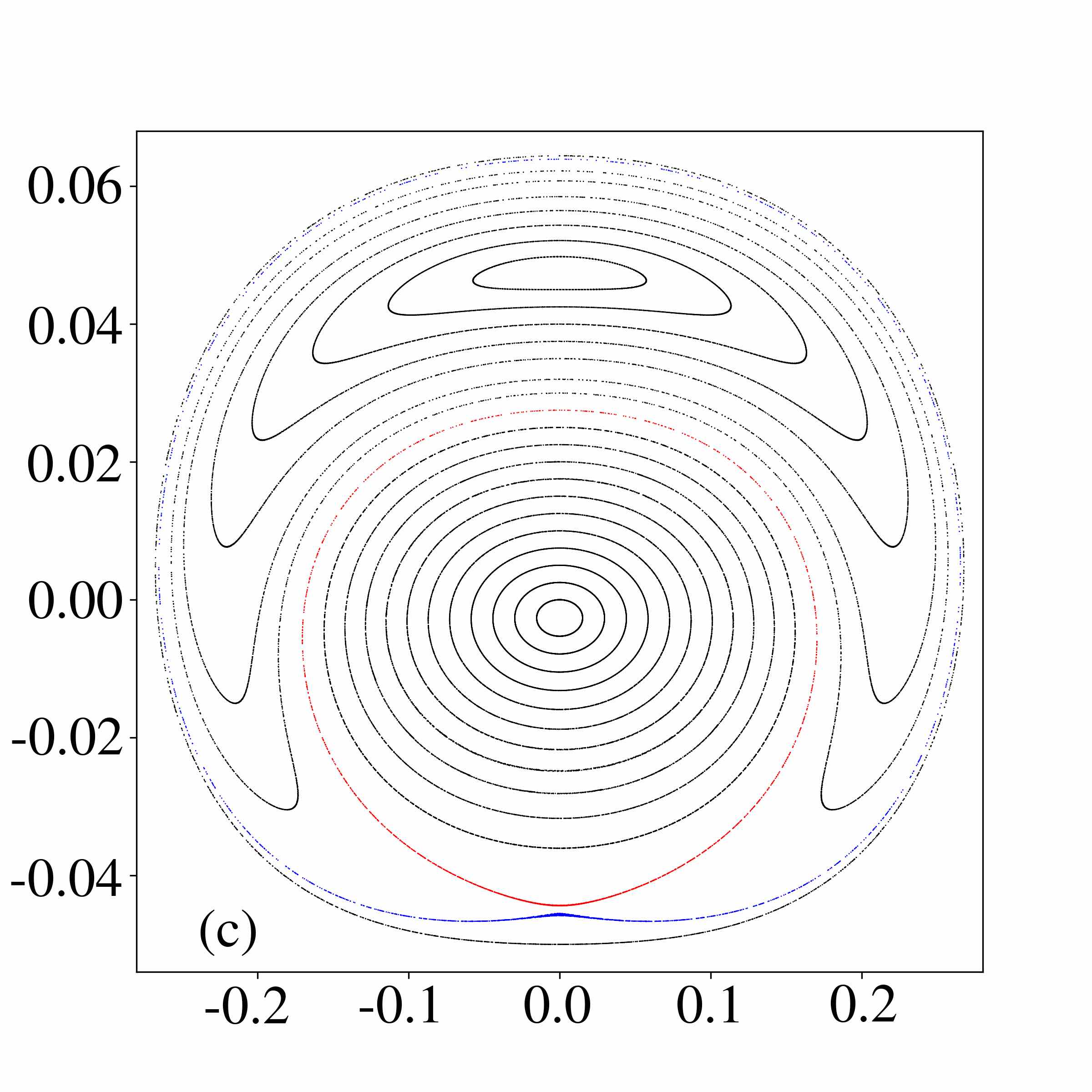}\\
\includegraphics[width=0.3\textwidth]{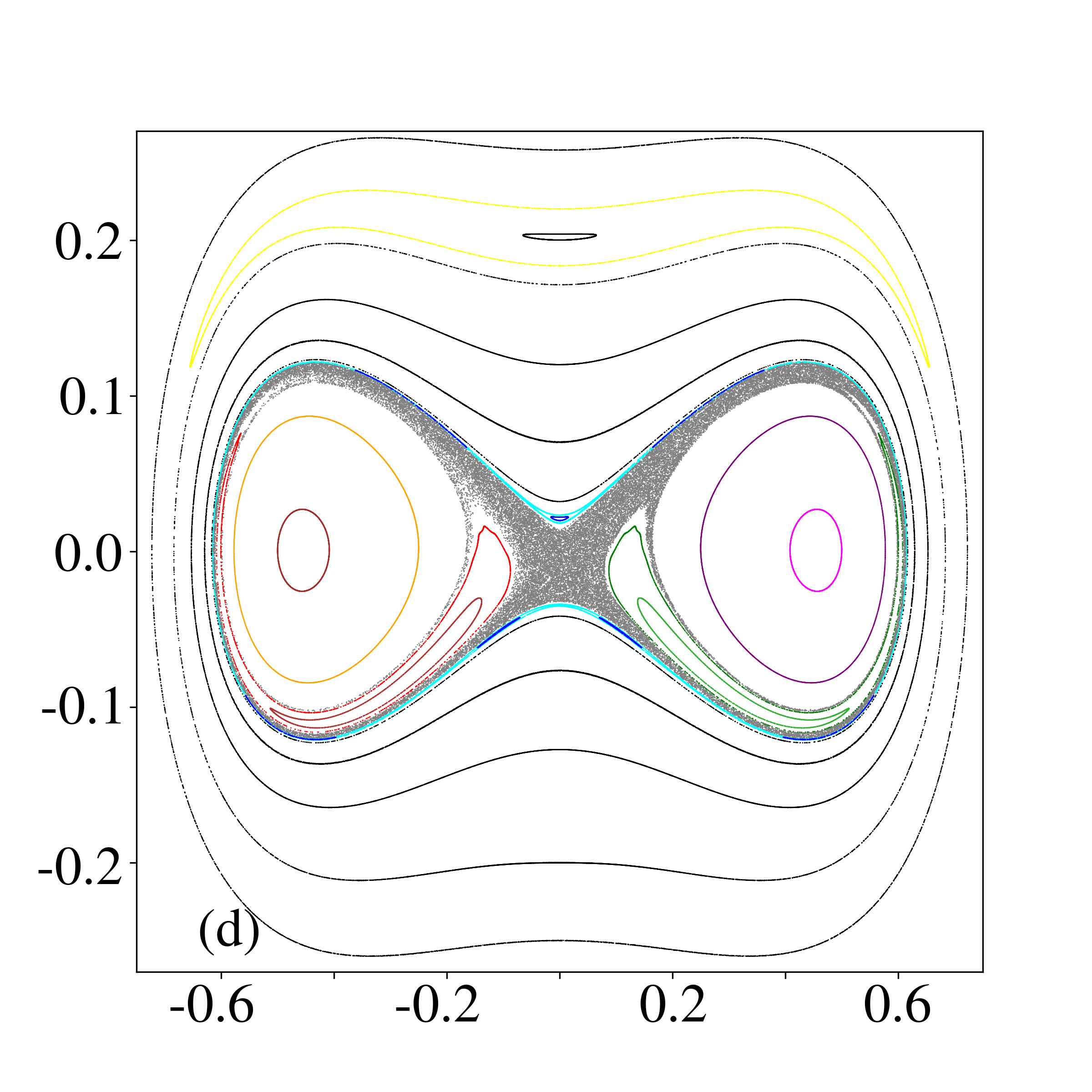}
\includegraphics[width=0.3\textwidth]{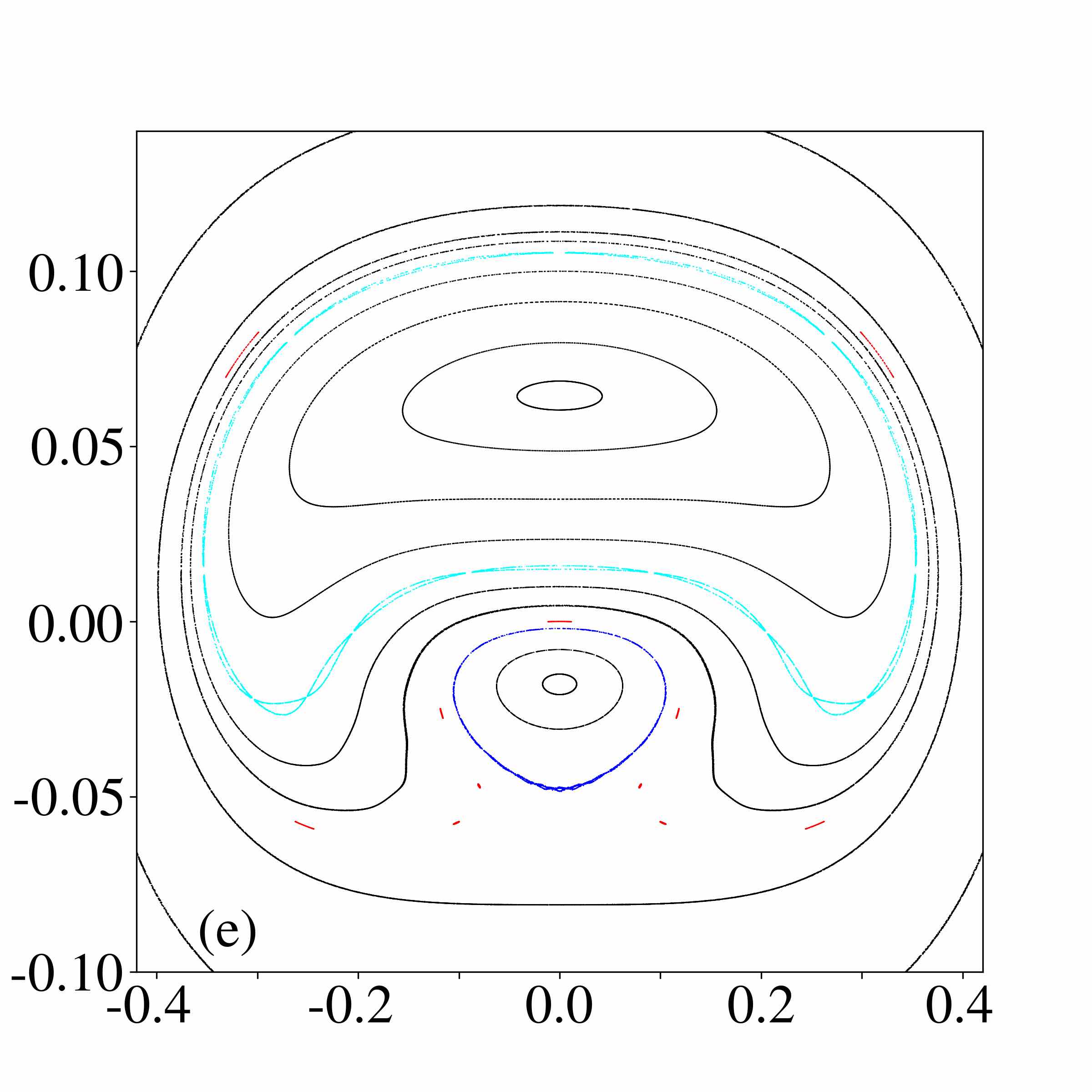}
\includegraphics[width=0.3\textwidth]{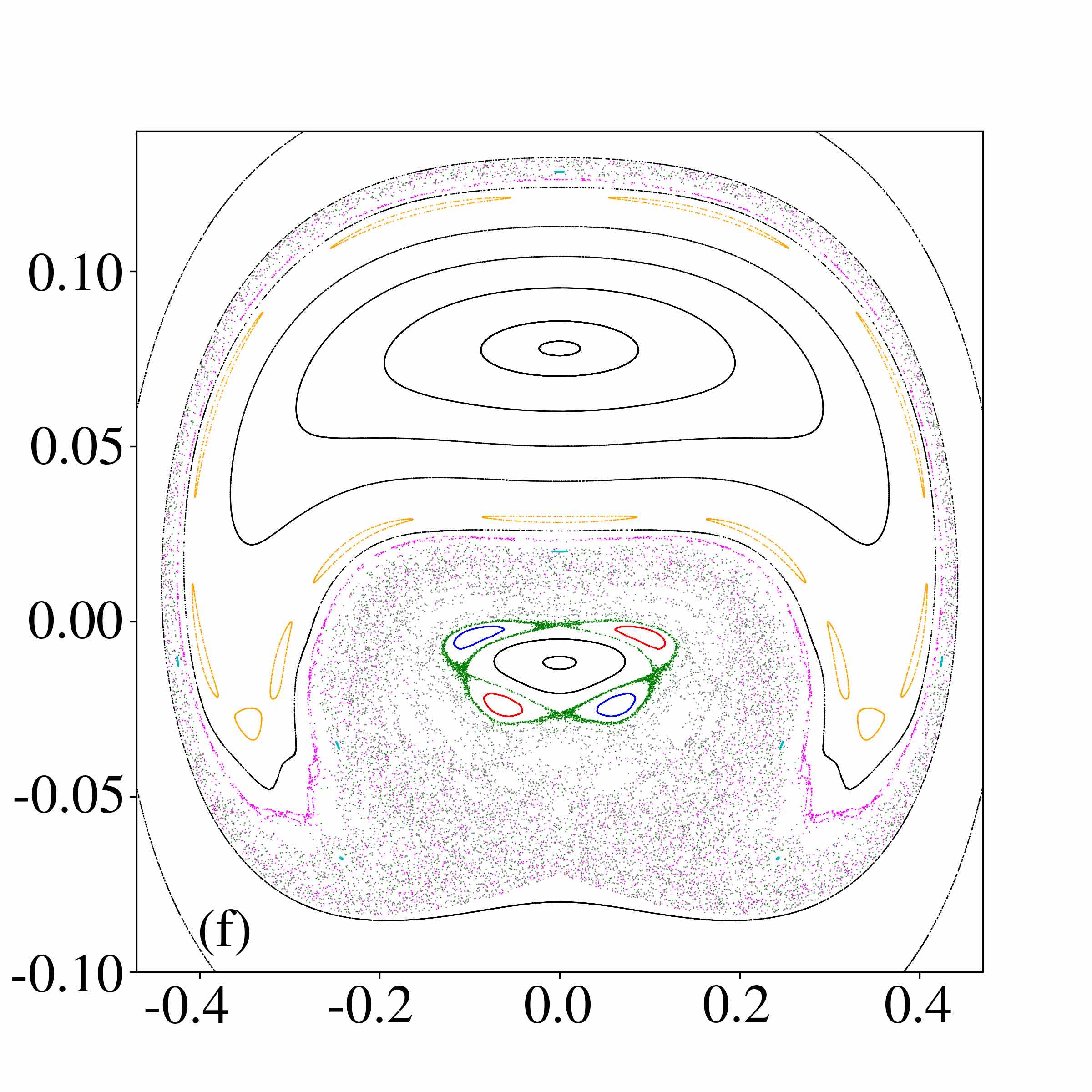}\\
\includegraphics[width=0.3\textwidth]{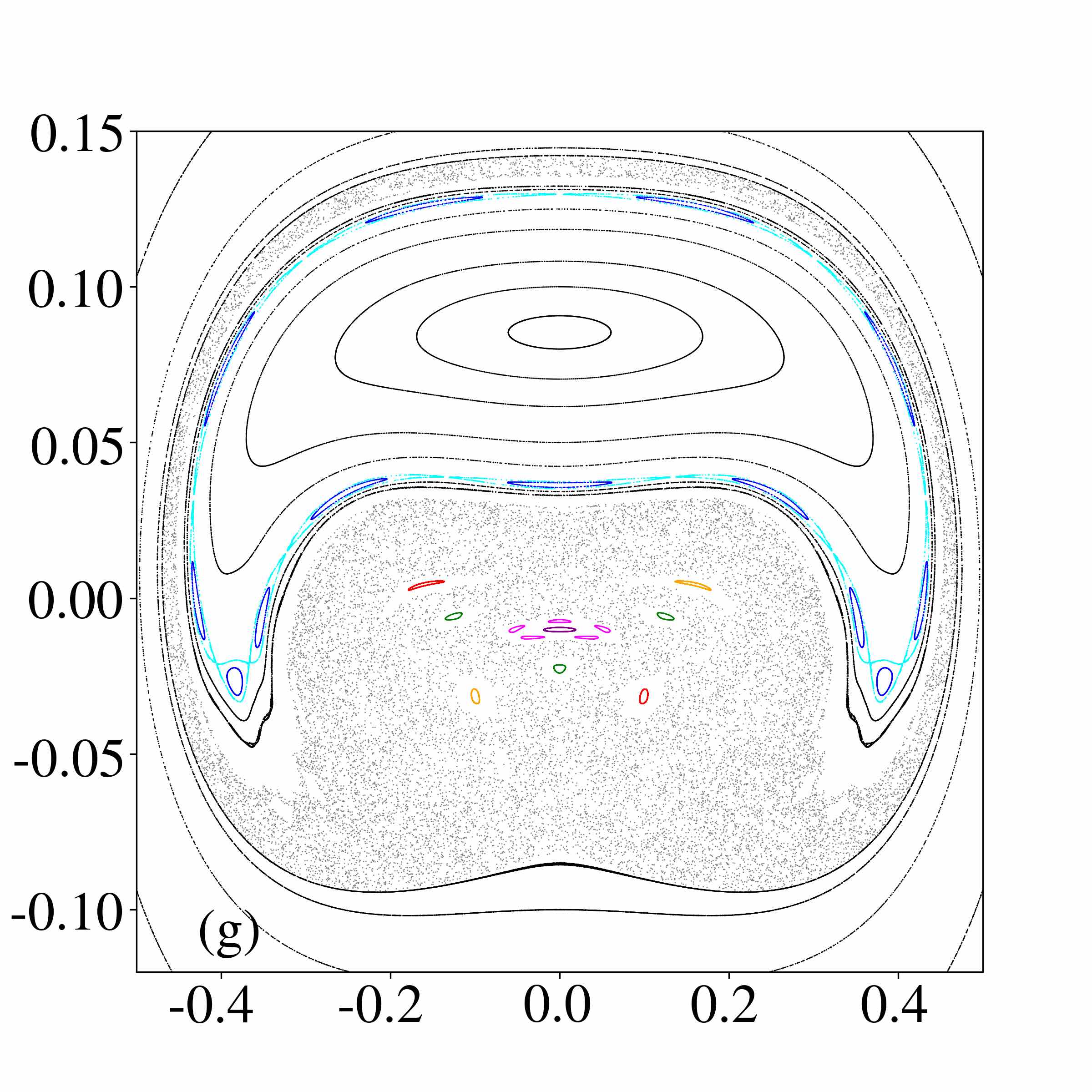}
\includegraphics[width=0.3\textwidth]{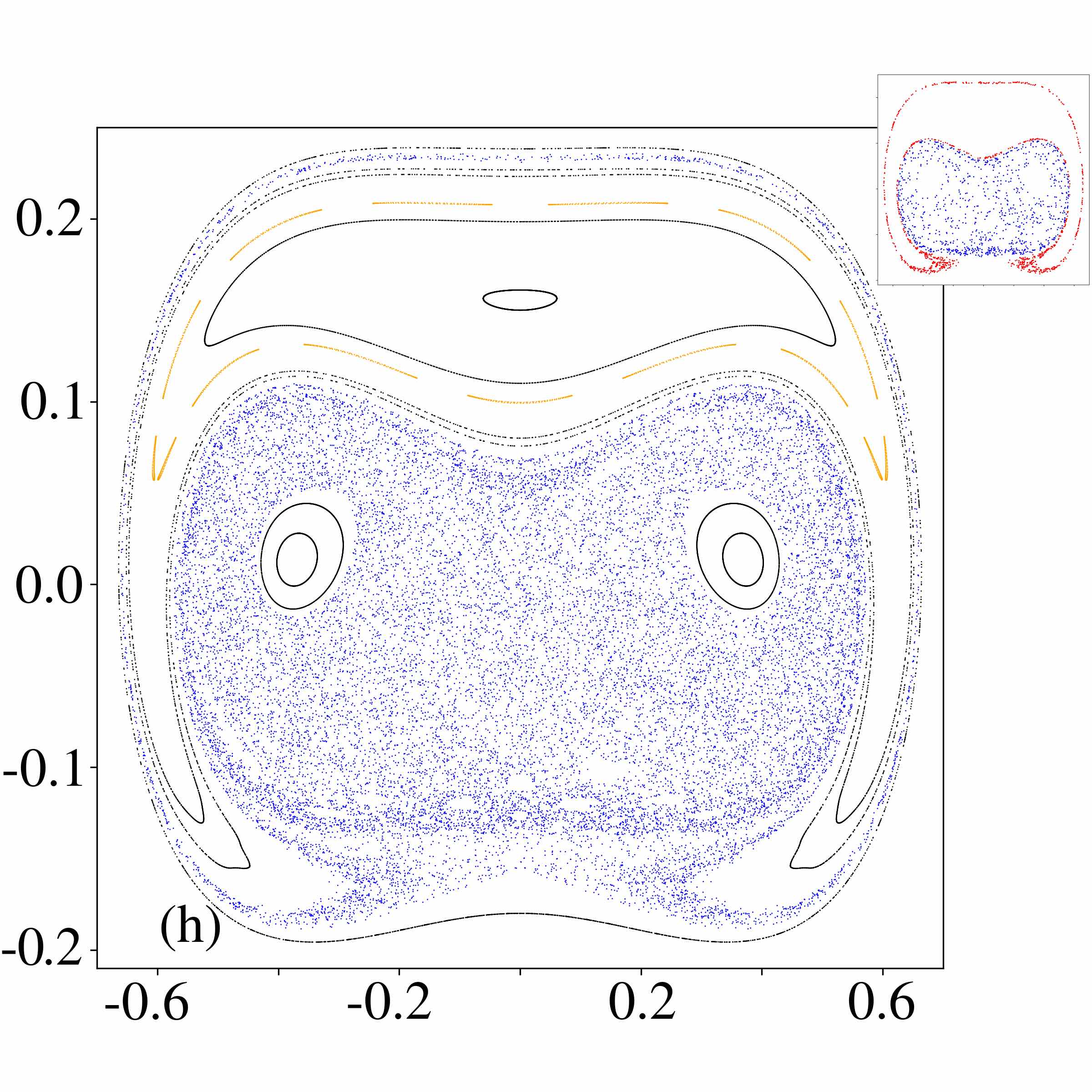}
\includegraphics[width=0.3\textwidth]{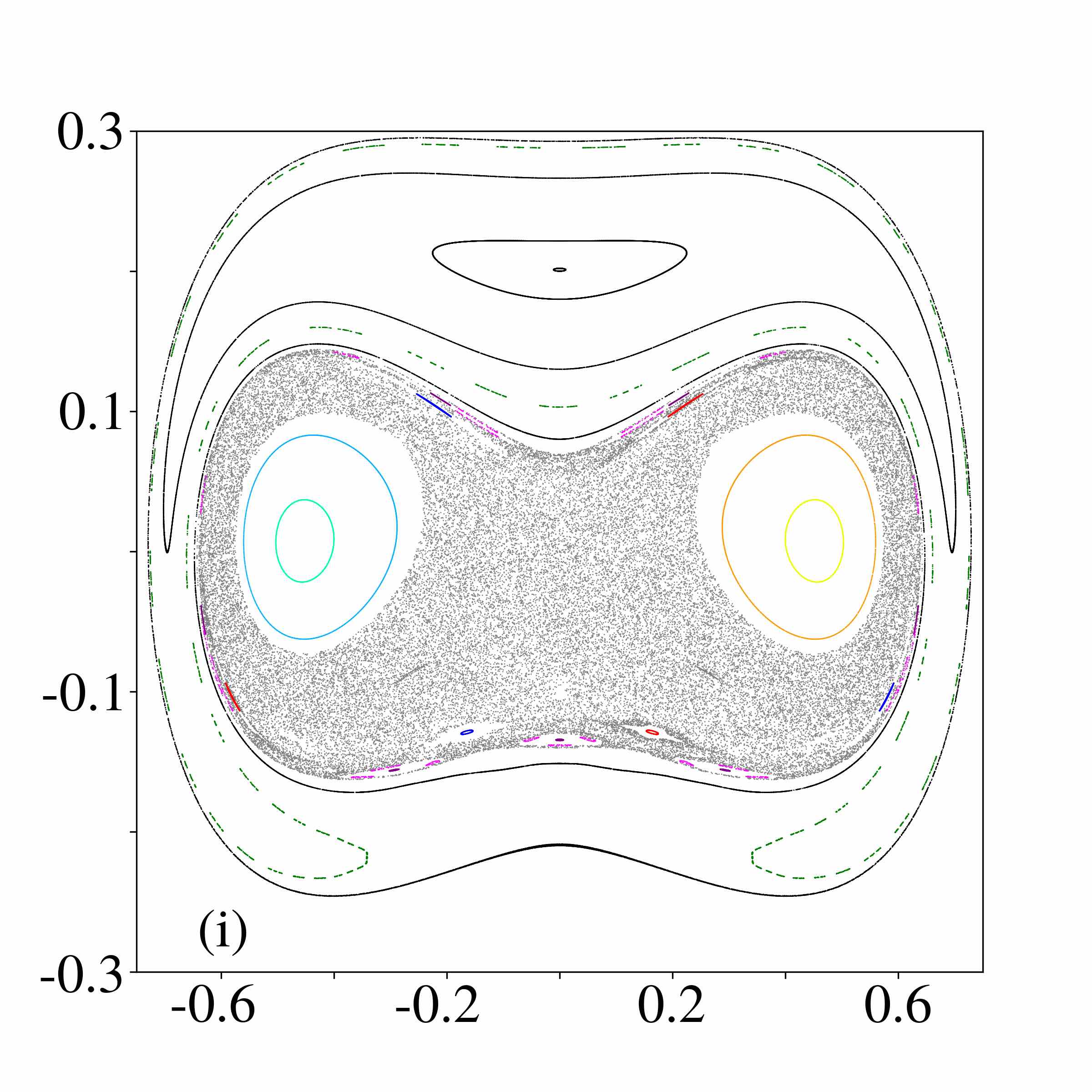}\\
\includegraphics[width=0.3\textwidth]{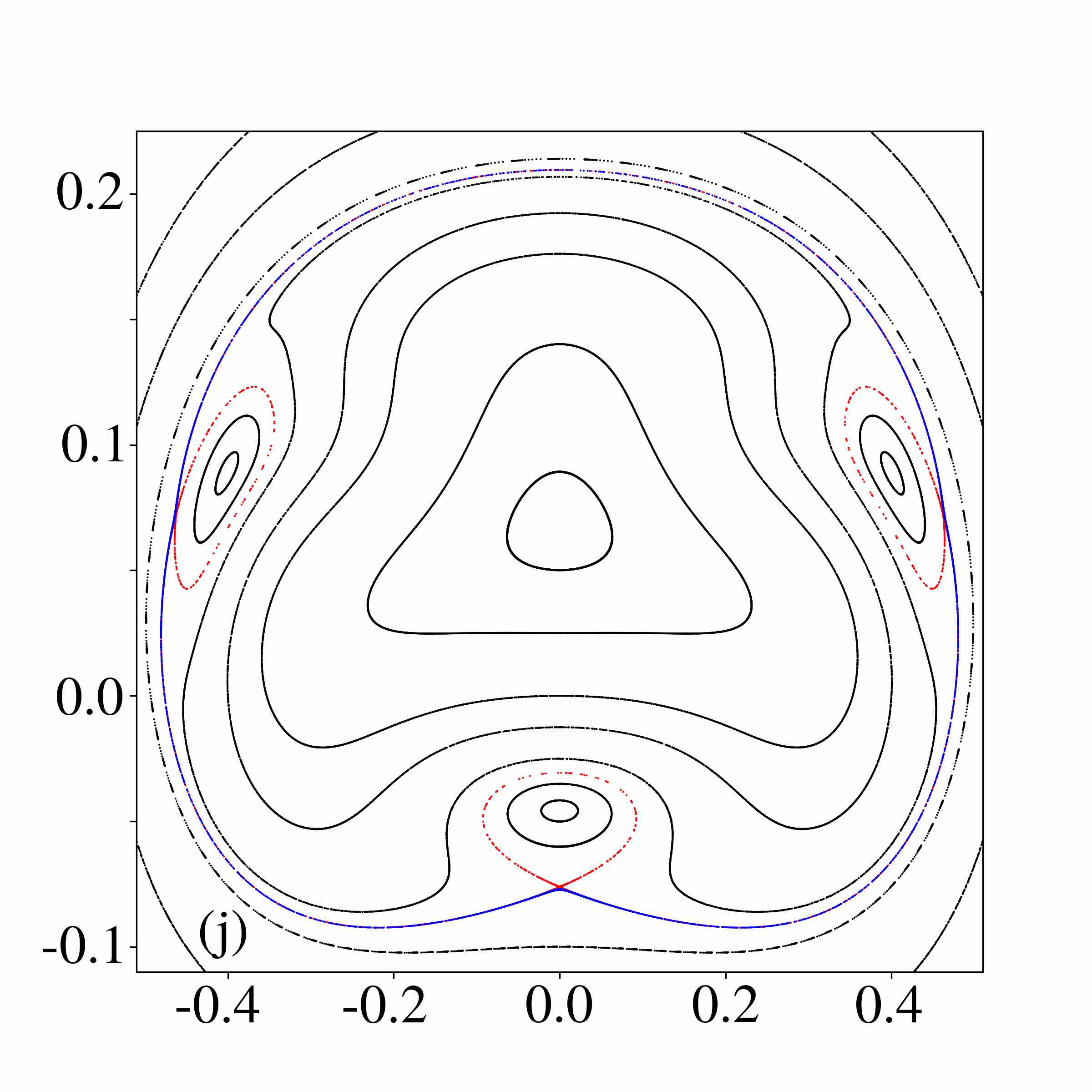}
\includegraphics[width=0.3\textwidth]{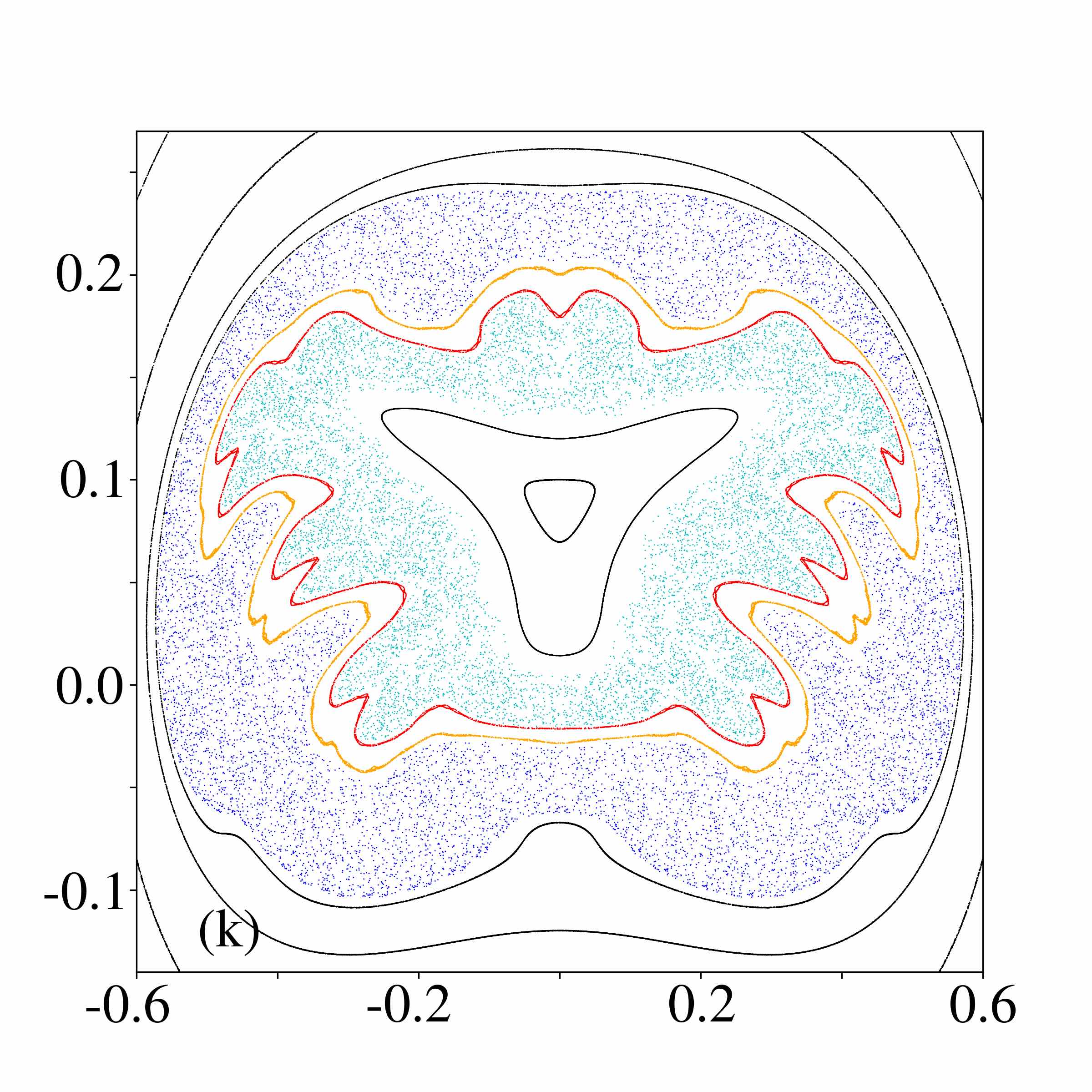}
\includegraphics[width=0.3\textwidth]{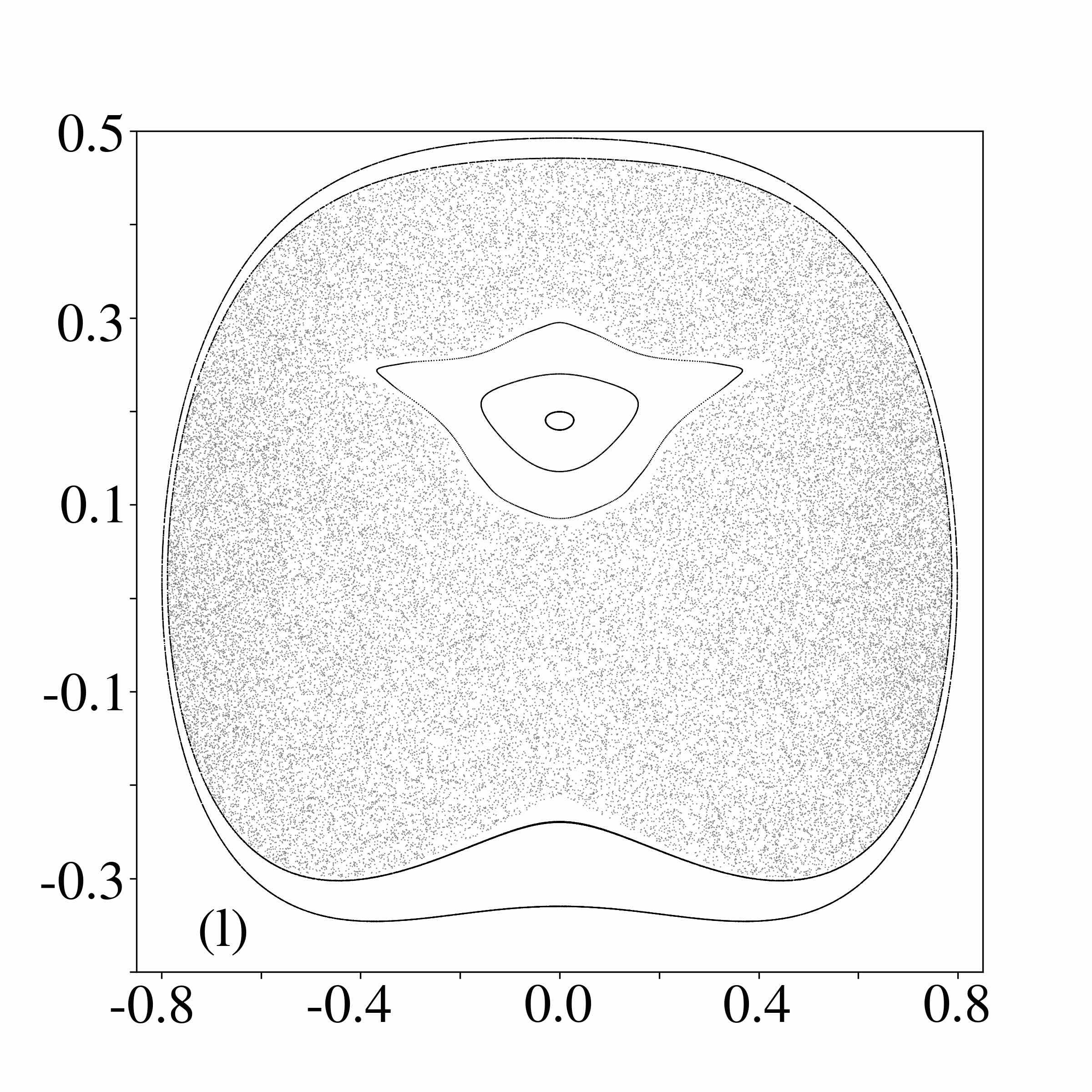}
\caption{\label{fig-ap:poincare}
Poincar\' e sections changing the parameters $h$ and $m$. Different trajectories are represented in different colors (except the ones in black, that do not highlight important features). (a)-(d) use $h = 10^{-3}$ with increasing $m$ (0.1, 0.2105, 0.22 and 0.4 respectively). (e)-(i) use $h = 10^{-2}$ with increasing $m$ (0.229, 0.242, 0.25, 0.34 and 0.4 respectively). (j)-(l) use $h = 10^{-1}$ with increasing $m$ (0.22, 0.25 and 0.4 respectively). The trajectories not presented are all quasi-periodic (reminiscent of the trajectories in (a)). A more detailed description can be found in the main text.
}
\end{figure*}

In this appendix we present a more detailed picture of how the chaotic solutions emerge in the presence of a forcing, using Poincar\'e sections (videos showing the change of the Poincar\'e sections as the chosen phase changes can also be found and can be useful to understand the time evolution along the phase space). The Poincar\'e sections themselves are in figure \ref{fig-ap:poincare}.

In \ref{fig-ap:poincare}a we see a situation with low $m$ and $h$ that is essentially the same as what we get without forcing ($h = 0$) below the critical mass ($m_c = 0.25$). In \ref{fig-ap:poincare}b, as $m$ increases, the curves start to deform until a cusp forms and a new family of solutions appears (in red). However since quasi-periodic solutions cannot cross with each other, this indicates the appearance of a chaotic solution separating the 2 families. In \ref{fig-ap:poincare}c, the solutions keep deforming. This keeps going until the homoclinic solution appears. Because of the forcing the homoclinic solution also becomes chaotic (\ref{fig-ap:poincare}d).

Increasing $h$, more complex structures start to appear. In \ref{fig-ap:poincare}e, the cyan and blue curves correspond to chaotic solutions, while the red one is evidence of another one, all of which separate different families of quasi-periodic solutions. These chaotic solutions start to merge as $m$ increases (\ref{fig-ap:poincare}f and \ref{fig-ap:poincare}g). The analog of the homoclinic solution is not as clear now (\ref{fig-ap:poincare}h), but eventually appears when $m$ becomes larger (\ref{fig-ap:poincare}i). An interesting detail is that this happens because the homoclinic solution detaches from the outermost chaotic solutions (a crisis). An evidence of this is that (\ref{fig-ap:poincare}h) displays intermittence. The inset shows 2 solutions for shorter times (red and blue), where the 2 regions trap the trajectories for a long time before switching to the other one. Another evidence of intermittence can be found in (\ref{fig-ap:poincare}f), where the magenta and green trajectories eventually get to the main chaotic region in gray (this can be seen magnifying the image). Since the dynamics is conservative, they should eventually return, but the time for that to happen is larger than the simulations we did.

As $h$ keeps increasing the vestiges of the unforced behaviour keep disappearing, including a completely different route to chaos (\ref{fig-ap:poincare}j to \ref{fig-ap:poincare}l).

\clearpage

\bibliographystyle{apsrev4-1}
\bibliography{library}

\end{document}